\documentclass[manuscript,screen]{acmart}
\usepackage{acronym}
\usepackage{braket}
\usepackage{enumitem}
\usepackage{listings,color}
\usepackage{subcaption}
\AtBeginDocument{%
  }

\setcopyright{acmcopyright}
\copyrightyear{2025}
\acmYear{2025}
\acmDOI{XXXXXXX.XXXXXXX}

\acrodef{API}{Application Programming Interface}
\acrodef{APIs}{Application Programming Interfaces}
\acrodef{BQC}{Blind Quantum Computation}
\acrodef{DFC}{Degree of Functional Corruption}
\acrodef{DNN}{Deep Neural Networks}
\acrodef{FGSM}{Fast Gradient Signing Method}
\acrodef{GHZ}{Greenberger-Horne-Zeilinger}
\acrodef{GPU}{Graphics Processing Unit}
\acrodef{IND-CPA}{Indistinguishability under Chosen Plaintext Attack}
\acrodef{IP}{Intellectual Property}
\acrodef{JSON}{JavaScript Object Notation}
\acrodef{MAC}{Multiply-Accumulate}
\acrodef{MQT}{Munich Quantum Toolkit}
\acrodef{NISQ}{Noisy Intermediate-Scale Quantum}
\acrodef{PGD}{Projected Gradient Descent}
\acrodef{PNI}{Probabilistic Noninterference}
\acrodef{QDK}{Quantum Development Kit}
\acrodef{QEC}{Quantum Error Correction}
\acrodef{QNN}{Quantum Neural Networks}
\acrodef{qOWF}{quantum-secure One-Way Functions}
\acrodef{QHE}{Quantum Homomorphic Encryption}
\acrodef{QFT}{Quantum Fourier Transform}
\acrodef{QPU}{Quantum Processing Unit}
\acrodef{QPUs}{Quantum Processing Units}
\acrodef{QPUs}{Quantum Processing Units}
\acrodef{RNG}{Random Number Generator}
\acrodef{TVD}{Total Variation Distance}
\acrodef{VQA}{Variational Quantum Algorithms}
\acrodef{VQE}{Variational Quantum Eigensolver}

\begin{document}

\title{Evaluating Security Properties in the Execution of Quantum Circuits}

\author{Paolo Bernardi}
\email{paolo.bernardi@phd.unipi.it}
\orcid{0009-0000-6668-591X}
\affiliation{%
  \institution{University of Pisa}
  \city{Pisa}
  \country{Italy}
}

\author{Giuseppe Bisicchia}
\email{giuseppe.bisicchia@phd.unipi.it}
\orcid{0000-0002-1187-8391}
\affiliation{%
  \institution{University of Pisa}
 \city{Pisa}
 \country{Italy}
}

\author{Gian-Luigi Ferrari}
\email{gian-luigi.ferrari@unipi.it}
\orcid{0000-0003-3548-5514}
\affiliation{%
  \institution{University of Pisa}
  \city{Pisa}
  \country{Italy}
}

\author{Antonio Brogi}
\email{antonio.brogi@unipi.it}
\orcid{0000-0003-2048-2468}
\affiliation{%
  \institution{University of Pisa}
  \city{Pisa}
  \country{Italy}
  }

\renewcommand{\shortauthors}{Bernardi et al.}

\begin{abstract}
Quantum computing is increasingly accessed through cloud platforms, where users delegate the execution of quantum circuits to remotely hosted NISQ devices owned by potentially untrusted providers. In this setting, both the confidentiality of circuits, input and output data and the integrity of the computation results may become primary concerns. This paper investigates whether circuit cutting, when suitably enhanced, can be used not only to overcome hardware size limitations but also as a practical security mechanism in cloud-based quantum computing. We introduce a set of integrity and confidentiality-oriented countermeasures, including dynamic QPU integrity scoring, probabilistic sub-circuit allocation, sub-circuit replication, and the injection of calibrated fake circuits. Guided by principles inspired by probabilistic noninterference, we define a heuristic evaluation framework for experimentally and quantitatively comparing the resilience of alternative configurations. Through an experimental campaign we evaluate how different countermeasure combinations influence the robustness of both global output correctness and input secrecy. Our findings show that enhanced circuit cutting can tolerate multiple malicious QPUs (up to 5 over 6 available QPUs) while preserving integrity, and that properly calibrated fake circuits can effectively obfuscate input circuits from adversarial observers.
\end{abstract}



\received{20 February 2007}
\received[revised]{12 March 2009}
\received[accepted]{5 June 2009}

\maketitle

\newpage
\section{Introduction}
\label{sec:intro}



Quantum computing is rapidly evolving, emphasising its potential across various domains, from cryptography~\cite{shor1995scheme} to molecular simulations~\cite{aspuru2005simulated}. Quantum computers are not yet widely available and are typically accessed remotely via cloud-based platforms, which provide computational resources such as quantum emulators, simulators, and experimental \ac{QPUs} over the internet~\cite{kaiiali2019cloud,rahaman2015review,bisicchia2024handcrafting,10313701,fi17110507}. 
This cloud-based access is not a limitation; rather, it facilitates effective experimentation and advances in quantum software development. Indeed, cloud-based access enables researchers to experiment with and validate techniques in preparation for more powerful quantum processors. Nonetheless, relying on third-party providers may expose quantum experiments to risks concerning the confidentiality and integrity of the processed information.

In this research context, innovative approaches such as \textit{quantum circuit cutting}~\cite{peng2020simulating,tang2021cutqc,bisicchia2024cut} offer promising ways to reduce computational demands. By dividing complex quantum circuits into smaller sub-circuits, this technique allows execution on hardware with limited capabilities. While quantum circuit cutting is widely explored for its performance and scalability benefits, there are, to the best of the authors' knowledge, only sparse references to its potential application as a security-enhancing method.

Addressing security concerns in cloud-based quantum computing begins with a clear understanding of the hostile environment and the associated threat model. In this paper, we contextualise the problem by assuming an adversary with both read and write access to (sub)circuits executed on cloud-connected QPUs. The attacker’s objective is either to infer confidential information or to deliberately alter the outcomes of the computation. To systematically reason about these risks, we adopt probabilistic non-interference~\cite{backes2004computational} as our driving methodological framework. This perspective enables us to formalise how information may leak through circuit interactions under adversarial influence, and to quantify the degree to which different constructions preserve security. Building on the idea of non-interference, we derive a practical heuristic for comparing the security properties of alternative quantum circuit-cutting variants, thereby identifying configurations that offer the strongest and most resilient protection against the modelled threats.

Probabilistic noninterference was originally introduced to formalise information-flow security in classical and probabilistic systems. However, despite its foundational importance, probabilistic noninterference is not directly applicable to quantum systems~\cite{6595825}, and this mismatch arises from several fundamental differences.

First, information in quantum systems is revealed exclusively through measurement, an operation that both disturbs the underlying state and can introduce interference between observers. This is in contrast to classical and probabilistic models, where observation is passive and does not alter the system being observed. As a result, the classical notions of “noninterference through observation” cannot be straightforwardly transferred to scenarios where the act of observing itself changes the available information.

Second, the evolution of quantum states is governed by unitary transformations, which have an algebraic structure fundamentally different from the stochastic matrices used in classical probabilistic systems. These differences affect how adversarial influence and compositional reasoning must be modelled. In particular, classical probabilistic noninterference assumes that processes interact through stochastic transitions on well-defined probability distributions, whereas quantum agents interact through linear operators over Hilbert spaces, where superposition and entanglement create correlations for which classical frameworks have no direct counterpart.

A further complication arises from entanglement, which deeply modify the boundary between local and global system behaviour: security properties cannot be evaluated purely locally, because certain activities on one subsystem may instantaneously affect the joint state. This intrinsically non-local structure challenges classical assumptions about how information flow can be constrained or partitioned.

Despite these limitations, probabilistic noninterference remains a valuable conceptual foundation. It provides an intuition for reasoning about how observable behaviour should depend on confidential inputs and offers a methodological starting point for constructing security metrics. In this paper, we apply this classical framework not as a strict correctness criterion, but as a guiding heuristic for comparing the relative security of quantum circuit-cutting strategies. By adapting its core idea, quantifying how adversarial observations probabilistically correlate with secret data, we design a  methodology for evaluating and contrasting different security configurations in cloud-based quantum computation.

The goals of this work are as follows:

\begin{enumerate}
\item to devise a practical methodology for ensuring integrity when executing quantum circuits within quantum cloud environments;
\item to devise a practical methodology for ensuring confidentiality  when executing quantum circuits within quantum cloud environments;
\item to introduce easy-to-apply heuristic measures for the rapid assessment of resilience against confidentiality and integrity breaches, enabling efficient and focused design iterations.
\end{enumerate}

In our research context and experimentation, probabilistic noninterference serves purely as an inspiration for the proposed heuristics. The formal verification of the circuit cutting-based methodology presented in the following sections is beyond the scope of this paper, and it is planned as future work.

The remainder of this paper is organised as follows. Section~\ref{sec:background} presents the basic principles of quantum circuit cutting. Section~\ref{sec:approach_overview} introduces the problem statement (architectural scenario, threat model, goals of the work) and a set of quantum circuit cutting security enhancements. Section~\ref{sec:exps} defines the heuristics to measure confidentiality and integrity resilience and applies them to experimentally assess the security enhancements configurations. Section~\ref{sec:threats} discusses potential threats to the validity of our findings. Section~\ref{sec:related} reviews relevant related work, while Section~\ref{sec:conclusions} concludes the manuscript with a critical discussion of our proposal.
\section{Background: Quantum Circuit Cutting}
\label{sec:background}

Quantum circuit cutting \cite{peng2020simulating} is a technique for the execution of complex quantum circuits over \ac{QPUs} with a limited number of input qubits, such as current \ac{NISQ}~\cite{Preskill:2018} devices. The core idea is to split a circuit into smaller sub-circuits with a reduced number of input qubits;  the sub-circuits can then be executed in parallel on the available \ac{QPUs}. 
To obtain the final result, the partial outputs from the QPUs must be reassembled with a classical post-processing step \cite{bisicchia2024cut}. Overall, this methodology makes it possible to expand the capabilities of current quantum devices at the cost of an exponentially complex reassembly procedure. Given a quantum circuit cut in $k$ fragments, the reassembly requires $O(16^k)$ time or $O(4^k)$ if the \ac{QPUs} can communicate with each other \cite{chen2023efficient}.

Circuit cutting is commonly performed by analysing and manipulating the circuit’s corresponding tensor network~\cite{Rudolph_2023}, a technique adapted from tensor-based quantum simulations. This method offers greater flexibility and a wider range of cutting strategies compared to straightforward qubit partitioning.

The cutting procedure involves a phase that can be summarised in three steps:
\begin{enumerate}
	\item the original quantum circuit is transformed in a tensor network;
	\item a network edge is cut, resulting in a collection of tensor networks;
	\item a set of smaller quantum sub-circuits is then derived from the tensor network.
\end{enumerate}

The cutting process can be applied iteratively to ensure that the number of qubits required by each sub-circuit aligns with the limitations of the available quantum hardware.

The resulting sub-circuits are distributed to a single \ac{QPU} one after the other (or to multiple \ac{QPUs} in parallel~\cite{bisicchia2024cut,tang2021cutqc}), and, finally, their partial outputs are collected.

We already remarked that the post-processing function constructs the global output through classical computation. This step is designed to produce an $\epsilon$ approximation of the tensor network result with a $2/3$ probability. The post-processing function is defined as follows:

\[
	f:\{0, 1\}^n \rightarrow [-1, 1].
\]

Each cut point in the circuit introduces a virtual qubit, and measurements are performed on multiple bases (such as Pauli bases). The post-processing function combines the results of each sub-circuit with the cut-point measurement outcomes. This algorithm often employs Kronecker products of matrices representing all possible sub-circuit outcomes at each cut point. These aggregated results are then employed to approximate the expectation value of the original quantum circuit.

As the number of cut points increases, so do the configurations to be considered since a new set of bases and their corresponding states are introduced at every cut. In the best case, this post-processing has a $O(4^k)$ time complexity, where $k$ is the number of cuts \cite{chen2023efficient}. This happens because the number of Kronecker products required is $4^k$.

Consequently, identifying optimal cut points is crucial, as the search space for a quantum circuit with 
$n$ edges can grow factorially, up to $O(n!)$. This makes the development of methods for automatic optimal cut-point selection an important area of research in the field  \cite{tang2021cutqc, ren2024hardware}.

Circuit cutting techniques have been shown to produce acceptable results on several quantum algorithms, ranging from Hamiltonian simulations performed with the \ac{VQE} algorithm \cite{peng2020simulating} to the Bernstein-Vazirani algorithm \cite{tang2021cutqc}.

In this paper, we argue that quantum circuit cutting can also serve as a \textit{security countermeasure}: intuitively, by strategically distributing circuit fragments and input data \textit{on multiple \ac{QPUs}}, no single \ac{QPU} is ever exposed to the complete circuit, the full input, or the entire computation output.
Here, to perform circuit cutting, we leverage the PennyLane framework by Xanadu \cite{bergholm2018pennylane}, which offers native support for differentiable quantum circuits and integrates with both software simulators and various \ac{QPUs}. Specifically, we use its built-in circuit cutting module, which automates the transformation into a tensor graph, the insertion of cut points, and the reconstruction of results. This enables us to focus our experiments on security without being encumbered by low-level implementation details.

\section{Overview of the Approach}
\label{sec:approach_overview}

Before presenting the full details of our approach, we first provide an informal overview to highlight its key features and underlying intuition. This high-level summary aims to offer readers a conceptual understanding of the main ideas guiding our methodology.

\subsection{Scenario}
\label{subsec:scenario}

Quantum computing providers offer \ac{APIs} that enable cloud-based access to their devices. In this work, we consider a scenario involving a centralised, trusted scheduler -- such as the Quantum Broker described in \cite{fi17110507,10.1007/978-3-031-48421-6_25} and \cite{10313701} -- which distributes quantum computations to individual \ac{QPUs} via classical network communication\cite{BISICCHIA2025108005}. Throughout the remainder of this paper, we refer to this setup as the \textit{cloud-based quantum computing scenario}.

The Quantum Broker can access several \ac{QPUs}, possibly operated by different vendors. For this reason, we consider a scenario where \ac{QPUs}, in general, cannot communicate with each other. This setup can be used in combination with Quantum Circuit Cutting to evaluate quantum circuits with a number of input qubits greater than the available \ac{NISQ} devices capabilities \cite{10.1007/978-3-031-48421-6_25}; moreover, distributing circuit fragments across multiple \ac{QPUs} enables parallel circuit execution.

In this paper, specifically, we consider the following workflow for the execution of a quantum circuit on multiple \ac{QPU}:
\begin{enumerate}
    \item the Broker splits the original circuit into multiple fragments with a Quantum Circuit Cutting algorithm;
    \item the Broker sends each fragment to a specific \ac{QPU};
    \item each \ac{QPU} computes the assigned fragment and produces a partial results;
    \item the Broker collects the partial results from each \ac{QPU};
    \item finally, the broker combines the partial results with a classical post-processing algorithm to obtain the result of the original circuit.
\end{enumerate}

Therefore, in this scenario, the more \ac{QPUs} available to the Quantum Broker, the higher the level of parallelism and flexibility. Using more \ac{QPUs} can decrease the overall execution time, although within limits. The complexity of the classical post-processing algorithm, in fact, grows with the number of circuit fragments: since we consider a scenario where inter-\ac{QPU} communication is not guaranteed, the upper bound of the classical post-processing algorithm time is $O(16^k)$, where $k$ is the number of circuit fragments.

In contrast, reducing the number of \ac{QPUs} available to the Quantum Broker decreases the level of parallelism and flexibility available while reducing the execution time of the post-processing algorithm.

Cybersecurity represents one of the most significant challenges in Quantum Software Engineering (we refer to \cite{murillo2024quantumsoftwareengineeringroadmap} for a more comprehensive discussion). This paper provides a novel contribution to the field of quantum software security by assessing and enhancing security methodologies that address information security requirements within a cloud-based quantum computing context.

\subsection{Threat Model}
\label{subsec:threat-model}

The scenario considered is characterised by a set of underlying security assumptions briefly explained below.

\begin{enumerate}
	\item There is a trusted central scheduler that distributes quantum circuits and input data to potentially untrusted \ac{QPUs} and collects their outputs;
	\item communication links within scheduler and \ac{QPUs} cannot be trusted;
	\item the \ac{QPUs} themselves are potentially untrustworthy.
\end{enumerate}

\begin{figure}[htbp]
	\begin{center}
		\includegraphics[width=0.8\textwidth]{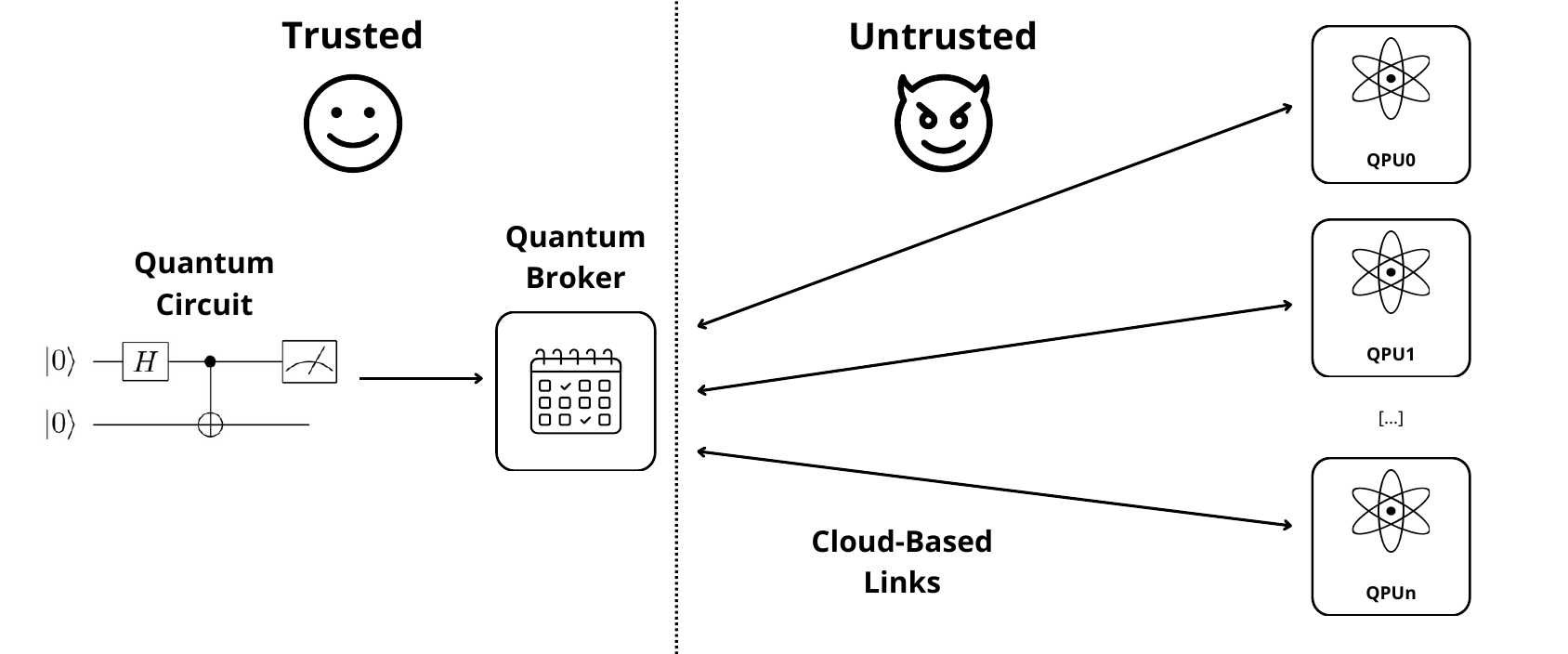}
		\caption{Graphical representation of the cloud-based quantum computing environment with the range of action of malicious actors.\label{fig:threat-model}}
        \Description{Graphical representation of the cloud-based quantum computing environment with the range of action of malicious actors.}
	\end{center}
\end{figure}

Figure~\ref{fig:threat-model} displays the environment and the trustworthiness of the various actors involved.

The potential objectives of the attackers constitute the basis of the threat model we consider:

\begin{enumerate}
	\item stealing designs for quantum circuits, as they could be \ac{IP} whose development required significant investments;
	\item stealing quantum computation results;
	\item disrupting quantum computation results, possibly without detection.
\end{enumerate}

The first two attacker objectives are related to quantum circuits and input data confidentiality, while the third revolves around computation output data integrity.

\subsection{Objectives}
\label{subsec:objectives}

Our work aims at achieving three main goals:

\begin{description}
	\item[Goal G1:] Given a trusted centralised scheduler and potentially untrusted quantum processors, we propose a methodology aimed at improving the \textit{integrity} of the output data.
	
	\item[Goal G2:] Given a trusted centralised scheduler and potentially untrusted quantum processors, we propose a methodology aimed at improving  the \textit{confidentiality} of algorithms and input/output data.

    \item[Goal G3:] Provide software engineers with a methodological framework to compare different quantum architectures with respect to integrity and confidentiality resilience.
\end{description}

\subsection{Quantum Circuit Cutting Security Enhancements}
\label{subsec:countermeasures}   

Although plain Quantum Circuit Cutting provides a baseline level of data segregation (as no single \ac{QPU} possesses all the circuit information), it remains vulnerable to the sophisticated adversaries defined in our threat model. To address this, we propose a  methodology that layers additional active countermeasures on top of the standard cutting procedure.

The countermeasures defined below come from intuitive considerations. In Section~\ref{sec:exps}, they will be evaluated with a quantitative methodology to experimentally determine local maxima with respect to integrity and confidentiality resilience.  

\subsubsection{Countermeasure i1 - Dynamic QPU integrity score (integrity protection)}

An integrity score is a numerical index that represents the reliability of a QPU with respect to integrity (e.g., the higher the score, the higher the QPU resilience against integrity attacks). Although this technique is not a security countermeasure per se, it is the cornerstone of the enhanced integrity protection proposed in this work.

The high-level process for computing integrity scores is as follows. 

\begin{enumerate}
    \item probe circuits with known expectation values are sent to \ac{QPUs};
    \item the difference between the \ac{QPUs} output and the ground truth is evaluated (see Equation \ref{eq:integrity-score});
    \item a normalised integrity score, inversely proportional to each \ac{QPU} difference is finally computed.
\end{enumerate}

\textbf{Rationale:} being able to compare how well each \ac{QPU} preserves integrity allows for targeted circuit fragment allocation.

\subsubsection{Countermeasure i2 - Probabilistic sub-circuit assignment (integrity protection)}

In order to maximise integrity resilience, sub-circuits are assigned to each \ac{QPU} in a probabilistic fashion, depending on its integrity score. In Subsection~\ref{subsec:integrity}, two probability computation criteria (proportional and exponential) are evaluated.

\textbf{Rationale:} Intuitively, allocating more circuit fragments to the most reliable \ac{QPUs}, while tending to avoid the least reliable ones, should enhance the overall resilience against integrity attacks.

\subsubsection{Countermeasure i3 - Sub-circuit replication (integrity protection)}

Sub-circuits are evaluated by more than one \ac{QPU}, depending on a \textit{replication factor}, and their results are averaged using the \ac{QPUs}'s integrity scores as weights.

\textbf{Rationale:} Having circuit fragments evaluated by more than one \ac{QPU}, and combining their results by weighing them according to their integrity score, can help mitigate the effect of unreliable \ac{QPUs}. This approach differs slightly from \textit{Probabilistic \ac{QPU} assignment}: rather than avoiding problematic \ac{QPUs}, it allows them to participate in the computation while reducing the impact of their results.

\subsubsection{Countermeasure c1 - Fake sub-circuits (confidentiality protection)}

Fake sub-circuits are mixed with real sub-circuits to make it harder for \ac{QPUs} to detect the actual computation being performed.

\textbf{Rationale:} mixing fake circuit fragments with real ones makes it harder for compromised \ac{QPUs} to infer sensitive data or reverse-engineer the original quantum circuit.

\begin{table}[ht]
    \centering
    \renewcommand{\arraystretch}{1.5}
    \setlength{\tabcolsep}{3pt}
    
    \begin{tabular}{ l l l }
        \hline
        \textbf{Enhancement} & \textbf{Description} & \textbf{Visual Representation} \\
        \hline
        
        \parbox[c]{0.28\textwidth}{\raggedright 
            \textbf{Integrity}: i1. Dynamic QPU integrity score
        } & 
        \parbox[c]{0.46\textwidth}{\raggedright 
            Probes \ac{QPUs} with known circuits to compute an \textit{integrity score} (Eq.~\ref{eq:integrity-score}) proportional to resilience.
        } & 
        \parbox[c]{0.20\textwidth}{
            \includegraphics[width=\linewidth]{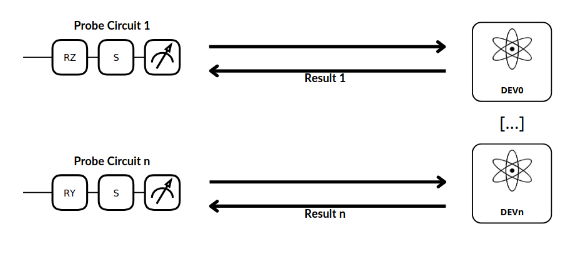}
        } \\
        \hline

        \parbox[c]{0.28\textwidth}{\raggedright 
            \textbf{Integrity}: i2. Probabilistic sub-circuit assignment
        } & 
        \parbox[c]{0.46\textwidth}{\raggedright 
            Assigns sub-circuits probabilistically; \ac{QPUs} with higher integrity scores receive a larger share of the workload.
        } & 
        \parbox[c]{0.20\textwidth}{
            \includegraphics[width=\linewidth]{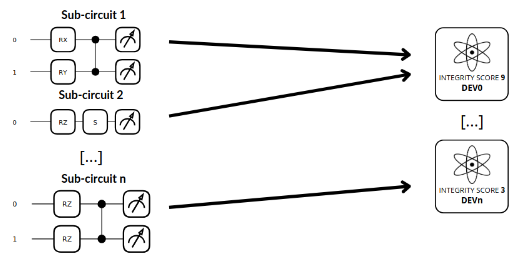}
        } \\
        \hline

        \parbox[c]{0.28\textwidth}{\raggedright 
            \textbf{Integrity}: i3. Sub-circuit replication
        } & 
        \parbox[c]{0.46\textwidth}{\raggedright 
            Replicates sub-circuits across multiple \ac{QPUs} (redundancy) and averages results weighted by integrity scores.
        } & 
        \parbox[c]{0.20\textwidth}{
            \includegraphics[width=\linewidth]{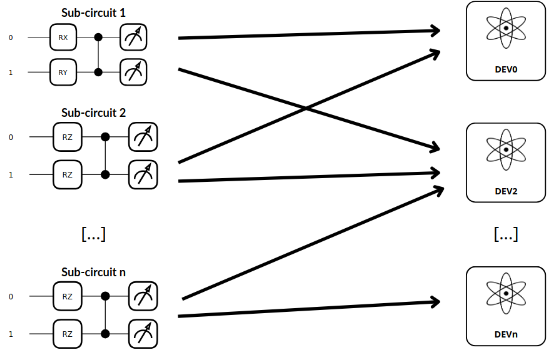}
        } \\
        \hline

        \parbox[c]{0.28\textwidth}{\raggedright 
            \textbf{Confidentiality}: c1. Fake sub-circuits
        } & 
        \parbox[c]{0.46\textwidth}{\raggedright 
            Mixes fake sub-circuits with real ones to obfuscate the computation and prevent reverse-engineering.
        } & 
        \parbox[c]{0.20\textwidth}{
            \includegraphics[width=\linewidth]{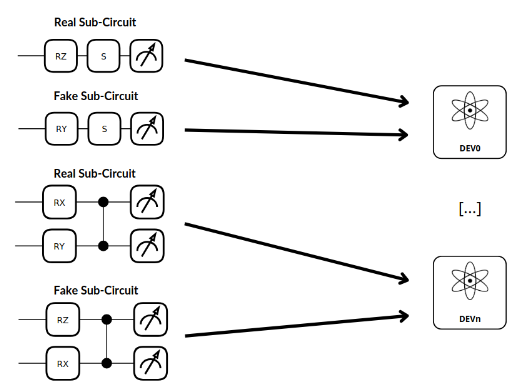}
        } \\
        \hline
    \end{tabular}
    \caption{Quantum Circuit Cutting security enhancements}
    \label{tab:countermeasures}
\end{table}

\clearpage

The integrity score, based on the results of the evaluation of the probe circuits by each \ac{QPU}, is as follows:

	\begin{align}
		error = expected\_result - actual\_result, \\
		integrity\_score = max(0, 10 - \dfrac{error}{expected\_result} \cdot 10).
        \label{eq:integrity-score}
	\end{align}

Each probe circuit is characterised by a known expectation value. To assess the integrity resilience of a \ac{QPU}, we firstly compute the difference between this theoretical value and the actual result produced by the device. This difference, referred to as the \textit{error}, reflects the deviation from the expected result: the higher the error, the less reliable the \ac{QPU} is considered. Finally, the \ac{QPU} integrity score is derived by normalising the error within a predefined range (in this example $0$ to $10$), so that a higher integrity score corresponds to greater \ac{QPU} reliability.

\section{Experimental Validation}
\label{sec:exps}

In this section, we illustrate the steps of the evaluation procedure designed to assess the extent to which the concept of circuit cutting can function as an effective security countermeasure against adversarial attacks, within the framework of cloud-based access to quantum computing devices.

We begin by outlining the key features of the testing methodology adopted in our approach. This includes a detailed description of the underlying assumptions: circuits, tools and test environment. We then move to illustrate the criteria used to assess the effectiveness of the security countermeasure. By clarifying these aspects, we aim to provide a transparent and reproducible basis for interpreting the obtained results.

\subsection{Setting the context}
\label{subsec:context_setup}

\subsubsection{Test Circuits}

The quantum circuits used in the experimental setting include standard algorithms, such as \ac{GHZ} and Deutsche-Jozsa, and custom quantum circuits specifically developed for the purposes of this evaluation activity. 

Firstly, standard quantum algorithms are employed to ensure relevance to practical real-world applications. Secondly, custom circuits with uniform structures and distinct gate patterns are employed to probe specific behavioural characteristics of quantum devices that may not be fully captured by standard circuits. To this end, our custom circuits incorporate seemingly random elements and regular, algorithmic structures. This dual approach enables a more comprehensive evaluation of the methodology across a broader range of quantum computational scenarios.

\begin{figure}[htbp]
	\begin{center}
		\includegraphics[width=0.4\textwidth]{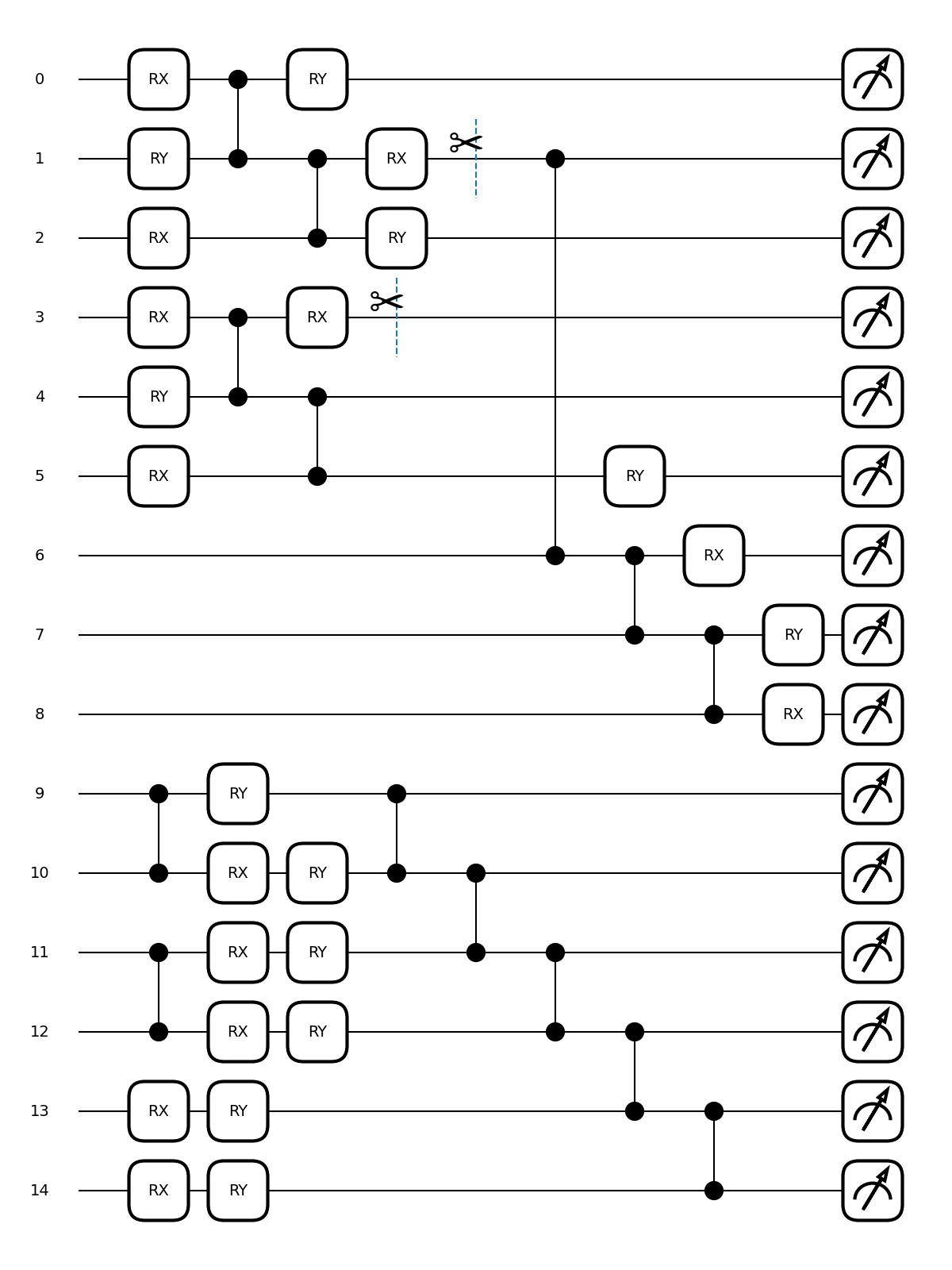}
		\caption{Graphical representation of the 15 qubit quantum circuit used to perform the experiments.\label{fig:benchmark-circuit}}
        \Description{Graphical representation of the 15 qubit quantum circuit used to perform the experiments.}
	\end{center}
\end{figure}

Our first custom benchmark circuit, displayed in Figure~\ref{fig:benchmark-circuit}, utilises 15 qubits and it is employed in integrity-resilience experiments. The first six qubits are initialised by combining rotations along the X and Y axes of the Bloch sphere through \lstinline|RX| and \lstinline|RY| gates. Each qubit is rotated by a specific angle.

Subsequently, qubits are linked with \lstinline|CZ| gates by applying a phase flip that depends on the respective control qubits. Then, the first four qubits are further processed with \lstinline|RX| and \lstinline|RY| gates. Finally, two cut points are added on wires 1 and 3 as a prerequisite for quantum circuit cutting. Up to the tenth, the remaining qubits are rotated and phase flipped using the \lstinline|RX|, \lstinline|RY|, and \lstinline|CZ| gates.

The gates that act on the remaining qubits are generated algorithmically. Each qubit (from the eleventh to the fifteenth) has a rotation \lstinline|RX| and a \lstinline|RY| proportional to its index, and it is phase-flipped by using the previous qubit as control.

The measurement consists of a \lstinline|Pauli-Z| on each qubit combined with a Kronecker product, as shown in equation \ref{eq_measurement}:

\begin{equation} \label{eq_measurement}
	Z_0 \otimes Z_1 \otimes Z_2 \otimes Z_3 \otimes \dots \otimes Z_{14}.
\end{equation}

\begin{figure}[htbp]
  \centering
  \begin{subfigure}[b]{0.48\textwidth}
    \centering
    \includegraphics[width=\linewidth]{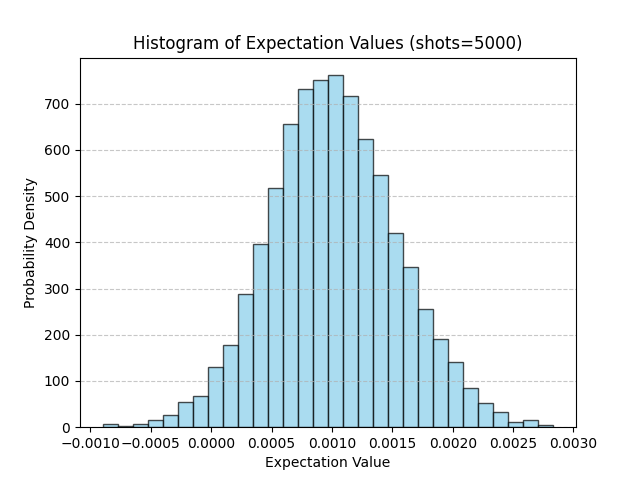}
    \caption{Histogram of the benchmark circuit expectation values}
    \Description{Histogram of the benchmark circuit expectation values}
    \label{fig:benchmark-circuit-histogram}
  \end{subfigure}\hfill
  \begin{subfigure}[b]{0.48\textwidth}
    \centering
    \includegraphics[width=\linewidth]{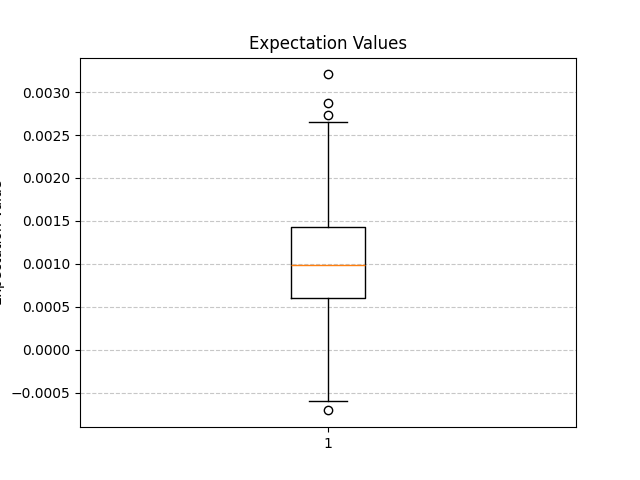}
    \caption{Box plot of the benchmark circuit expectation values}
    \Description{Box plot of the benchmark circuit expectation values}
    \label{fig:benchmark-circuit-boxplot}
  \end{subfigure}
  \caption{Graphical analysis of the benchmark circuit expectation values}
  \label{fig:benchmark-analysis}
\end{figure}

The expectation values for the benchmark circuit follow a Gaussian distribution described by the parameters $\mu = 0.001013413$ and $\sigma = 0.000000282$. Figure~\ref{fig:benchmark-circuit-histogram} shows the histogram of the expectation values derived with an experiment of 5000 circuit evaluations (5000 shots each), while Figure~\ref{fig:benchmark-circuit-boxplot} displays the box plot diagram of the expectation values.

Furthermore, we designed a second custom circuit for confidentiality testing, referred as the alternative benchmark circuit. Compared to the first one, it is structurally simpler but it has a larger percentage of noisy gates.

\begin{figure}[t]
	\begin{center}
		\includegraphics[width=0.3\textwidth]{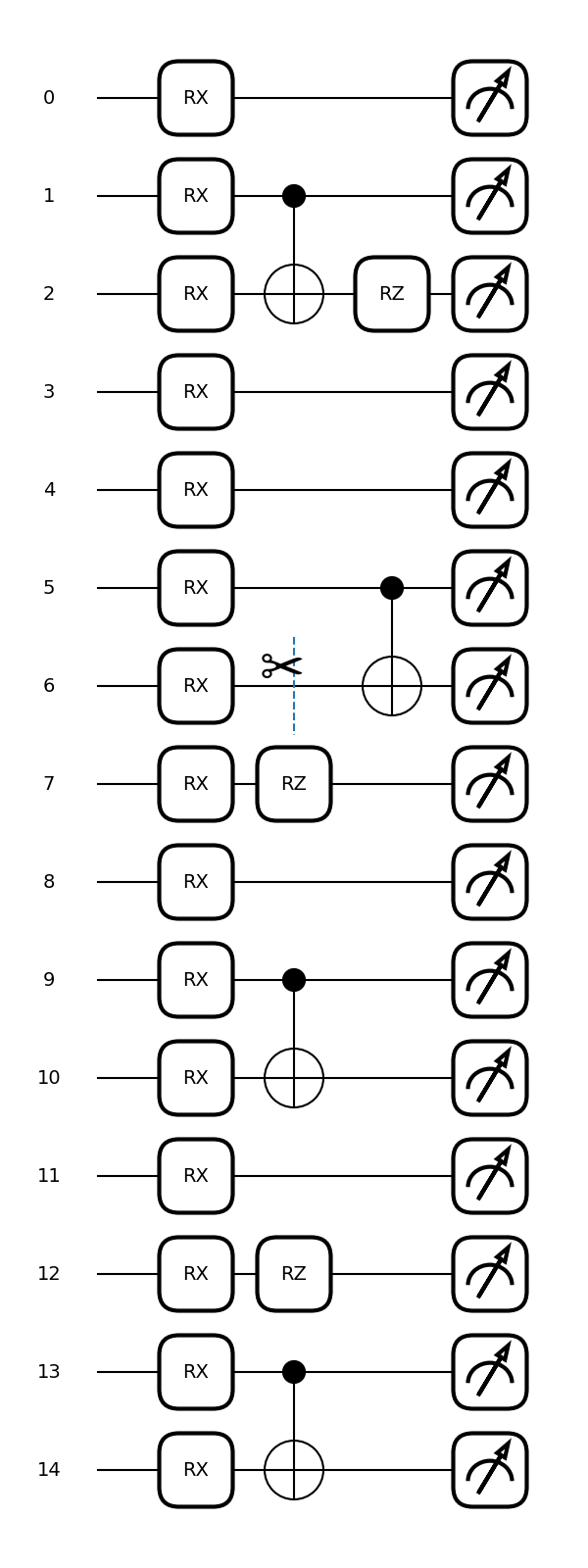}
		\caption{Graphical representation of the alternative 15 qubit quantum circuit used to perform confidentiality experiments together with the original benchmark circuit, \ac{GHZ} and Deutsch-Jozsa \label{fig:benchmark-circuit-alt}}
        \Description{Graphical representation of the alternative 15 qubit quantum circuit used to perform confidentiality experiments together with the original benchmark circuit, \ac{GHZ} and Deutsch-Jozsa}
	\end{center}
\end{figure}

Figure~\ref{fig:benchmark-circuit-alt} shows the second custom circuit structure. While the gates are different from the first custom benchmark circuit, it is still a 15 qubit circuit, and the measurements consist of \lstinline|Pauli-Z| observables on each qubit combined with a Kronecker product as well:

\begin{equation} \label{eq_measurement_alt}
	Z_0 \otimes Z_1 \otimes Z_2 \otimes Z_3 \otimes \dots \otimes Z_{14}.
\end{equation}

\begin{figure}[htbp]
    \centering
    \begin{subfigure}{0.48\textwidth}
        \centering
        \includegraphics[width=\linewidth]{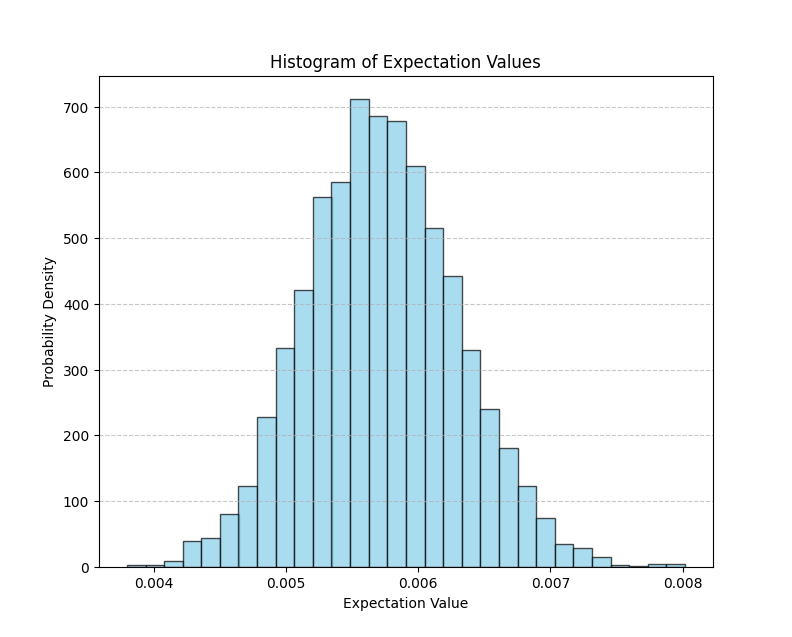}
        \caption{Histogram of the alternative benchmark circuit expectation values}
        \label{fig:benchmark-circuit-histogram-alt}
    \end{subfigure}
    \hfill
    \begin{subfigure}{0.48\textwidth}
        \centering
        \includegraphics[width=\linewidth]{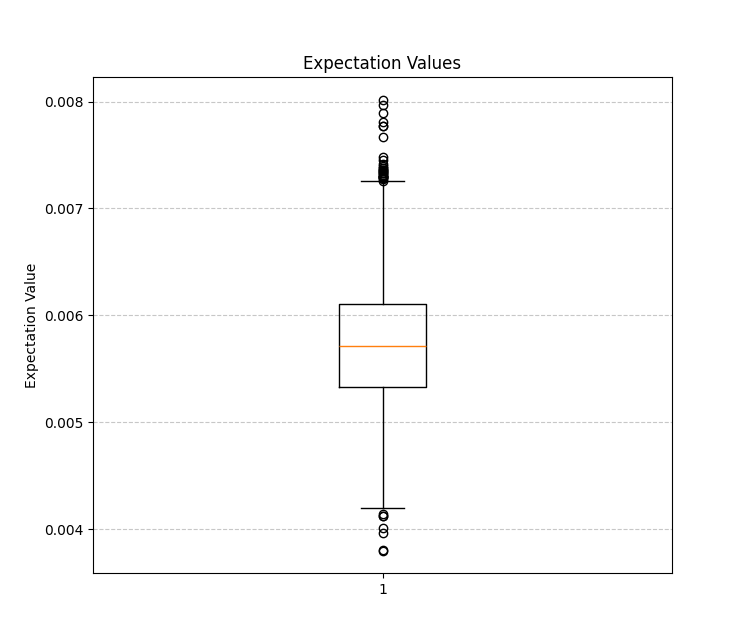}
        \caption{Box plot of the alternative benchmark circuit expectation values}
        \label{fig:benchmark-circuit-boxplot-alt}
    \end{subfigure}
    
    \caption{Graphical analysis of the alternative benchmark circuit expectation values}
    \label{fig:benchmark-circuit-analysis}
    \Description{Graphical analysis of the alternative benchmark circuit expectation values}
\end{figure}

The alternative benchmark circuit expectation values follow a Gaussian distribution described by the parameters $\mu = 0.005726323$ and $\sigma = 0.000000327$. Figure~\ref{fig:benchmark-circuit-histogram-alt} shows the histogram of the expectation values derived via an experiment of 5000 circuit evaluations with 5000 shots each, while Figure~\ref{fig:benchmark-circuit-boxplot-alt} displays the box plot diagram of the expectation values.

\subsubsection{Test Devices}

The availability of cloud-accessible \ac{NISQ} quantum processors from commercial vendors marks a significant milestone in the development and practical understanding of quantum computing. These quantum devices can be accessed through software libraries such as Qiskit \cite{qiskit2024}, which interface with the hardware via a custom token-based HTTP API. However, current \ac{QPUs} are not yet practical in a context where running hundreds or thousands of tests is required, since free-tiers have significant queue times, strict usage time accounting and non-free hardware access is considerably expensive.

Therefore, the following experiments have been performed on software-based noisy \ac{QPU} simulators\footnote{The code used to run the experiments and the experimental results are available at \url{https://github.com/pbswengineering/secprop-qcircuit-exec} and at \url{https://zenodo.org/records/17977895}.}. The test \ac{QPUs} used in this work consist of six Qiskit Aer simulators. The flexibility of Qiskit Aer made it possible to perform experiments both Intel CPUs and NVIDIA GPUs. The devices were configured to support 15 qubits, matching the input size of the circuits used in the experiments. This specific input size has been chosen as a trade-off between meaningful computational input and test execution time.

Depending on the specific testing requirements, the test devices  run in two different ways:

\begin{enumerate}
	\item \textbf{Reliable}: the simulated \ac{QPU} returns the actual \textit{true} computation result.
	
	\item \textbf{Malicious}: the \ac{QPU} alters the output result up to $250\%$. The actual magnitude of the attack is defined probabilistically through a \ac{RNG} with a configurable seed value. Specifically, given the $i^{\text{th}}$ result of the $n^{\text{th}}$ \ac{QPU}, the result is altered as follows:
	\[
	res_{n,i} = (1.5 + r) res_{n,i},
	\]
	where $r$ is a random number in the $[0,1)$ range.
\end{enumerate}

\subsubsection{Noise Profile}
\label{subsec:noise-profile}

The noise profile used for the experiments involves only three gates: $RX$, $RY$, and $CNOT$. Their behaviour is altered as follows:

\begin{itemize}
	\item $RX$ and $RY$ gates are subject to thermal relaxation errors with 2\% and 1.5\% probability respectively. This error simulates the decoherence caused by the interaction with the environment, given by the parameters $T_1$ ($50 \mu s$, the average relaxation time toward the $\ket{0}$ state) and $T_2$ ($30 \mu s$, the average dephasing time). The density matrix evolution is characterised by a combination of $T_1$ exponential decay and $T_2$ decoherence.
	\item The $CNOT$ gate is subject to depolarisation, a random noise that affects the whole quantum state space. This error simulates hardware errors due to qubit instability or external interference. The following formula describes the depolarisation error effects on the density matrix: 
\[
\mathcal{E}(\rho) = (1 - p) \rho + \frac{p}{d} I,
\]
where $\rho$ is the initial density matrix, $p$ is the depolarisation probability, $d$ is the system size (e.g. $d = 4$ for a two-qubit system), and $I$ is the identity matrix.
\end{itemize}

\begin{figure}
	\begin{lstlisting}[caption={\textit{Noise model}},label={cod:noise-model},language=Python]
cnot_error = depolarizing_error(0.01, 2)
rx_error = thermal_relaxation_error(50e-6, 30e-6, 0.02)
ry_error = thermal_relaxation_error(50e-6, 30e-6, 0.015)
noise_model.add_all_qubit_quantum_error(cnot_error, ['cx'])
noise_model.add_all_qubit_quantum_error(rx_error, ['rx'])
noise_model.add_all_qubit_quantum_error(ry_error, ['ry'])
	\end{lstlisting}
    \Description{Noise model definition}
\end{figure}

The Qiskit API exposes simulated devices, such as \lstinline|Fake127QPulseV1|, a 127 qubit processor with a complex topology inspired by IBM's Eagle \ac{QPUs}. This model allows for fine-grained parameter control for pulse-level simulation, including pulse duration and shape. Additional customisable parameters decoherence times for each qubit, reading error probability, gate error rates, and inter-qubit links.

In our tests we applied the noise profile defined above to a \lstinline|Fake127QPulseV1| simulated device. This specific device was selected since it mimics the topology of real-world devices, while the noise profile was defined to ensure controlled and interpretable evaluation conditions. The selected gates, $RX$, $RY$, and $CNOT$, are the most common gates used in our testing circuits. Applying thermal relaxation to single-qubit rotations reflects realistic decoherence phenomena common in superconducting quantum hardware. The depolarising noise applied to the $CNOT$ gate, instead, captures the probabilistic nature of multi-qubit gate failures, which are typically the most error-prone operations in \ac{NISQ} devices. This specific noise setup provides a simplified, easy to interpret, yet meaningful abstraction of typical noise sources.

\subsubsection{Comparison of Quantum Circuit Evaluation Results}

The experiments to validate integrity and confidentiality resilience of Quantum Circuit Cutting-based schemes depend crucially on comparing the output probability distributions of the benchmark quantum circuits.

To compare probability distributions we employ Hellinger Distance, as it is customary in the quantum computing field \cite{dasgupta2021stability, belovs2019quantum, luo2004informational, pitrik2020quantum}:

\begin{equation} \label{eq:hellinger}
	H(P, Q) = \frac{1}{\sqrt{2}} \sqrt{\sum_{x \in \mathcal{X}} \left( \sqrt{P(x)} - \sqrt{Q(x)} \right)^2},
\end{equation}
Hellinger Distance varies from 0 (the probability distributions $P$ and $Q$ are exactly the same) to 1 (they are completely different).

In the experimental results analysis, we will consider two distributions to be sufficiently similar if their Hellinger distance is at most 0.25. This threshold was determined considering two factors:

\begin{enumerate}
    \item \textbf{Distribution overlap:} considering the relationship between Hellinger distance and \ac{TVD} defined above, for a Hellinger distance threshold of 0.25, the \ac{TVD} is at most 0.35, which means that almost two thirds of the distributions do not overlap.
    \item \textbf{Bayes error probability} 
for a Hellinger distance threshold of 0.25, the corresponding Bayes error probability is approximately 1.5\%. Therefore, a classifier trying to distinguish between two distributions having this threshold would have a considerably low error probability.
\end{enumerate}

\subsection{Integrity Testing}
\label{subsec:integrity}

In order to rigorously assess the effectiveness of the proposed integrity countermeasures, we conducted a systematic experimental campaign. The primary objective was to quantify the resilience of the quantum circuit cutting process against active integrity attacks and to determine the maximum number of tolerated attackers for each protection scheme.

The experimental workflow determines and isolates incrementally the contribution of each countermeasure. We first establish a baseline using raw, unprotected circuit cutting. Subsequently, we evaluate the proposed enhancements in increasing order of complexity:  proportional probabilistic assignment with sub-circuit replication, exponential probabilistic assignment with sub-circuit replication, and exponential assignment without sub-circuit replication. For every configuration, we simulate a hostile environment by progressively increasing the number of malicious \ac{QPUs}, from a single attacker up to a fully compromised back-end, and measuring the deviation of the output distribution from the ground truth.

In order to quantify the degree of integrity protection guaranteed by Quantum Circuit Cutting, and to compare the effectiveness of the additional countermeasures defined in Subsection~\ref{subsec:countermeasures}, we define an heuristic framework inspired by \ac{PNI}.

We begin by dividing output data in two categories:
\begin{description}
	\item $O_{hi}$ is the high-level output that malicious \ac{QPUs} are trying to subvert;
	\item $O_{lo}$ is the low-level output produced by well behaving \ac{QPUs};
    \item $O_{lo'}$ is the low-level output produced by  malicious \ac{QPUs};
\end{description}

If an integrity-preserving procedure guarantees integrity, then the following probabilistic relationship should be verified:
\begin{equation} \label{eq:pni-integrity2}
	P(O_{hi} | O_{lo}) \simeq P(O_{hi} | O_{lo'}),
\end{equation}
This means that partial outputs from malicious \ac{QPUs} should not alter the overall computation output.

In our experimental assessment, we employed the following procedure to practically verify the validity of Equation~\ref{eq:pni-integrity2}:

\begin{enumerate}
	\item Compute the probability distribution of the benchmark circuit output (\textit{ground truth});
	\item Compute the Hellinger distance between the ground truth and the distribution produced by a system where 1 \ac{QPU} is acting maliciously, by altering its output $O_{lo}$;
	\item Add one malicious \ac{QPU} (so now there are two attackers) and measure the distance from ground truth again;
	\item Repeat step 3 until all 6 \ac{QPUs} are acting maliciously.
\end{enumerate}

As specified in subsection \ref{subsec:context_setup}, attacking \ac{QPUs} can alter their $O_{lo}$ values by up to 250\%.

In the following sections, guided by the principles of PNI we test combinations of the proposed security enhancement techniques on different circuits and with an increasing number of malicious \ac{QPUs} (i.e. saboteurs) and measure how they improve the integrity of the overall quantum computation.

\subsubsection{First test: Raw Circuit Cutting}

Firstly, the impact of attacking \ac{QPUs} was evaluated over plain quantum circuit cutting without any additional protection.

\begin{figure}[htbp]
	\centering
	\begin{subfigure}{0.45\textwidth}
		\centering
		\includegraphics[width=\textwidth]{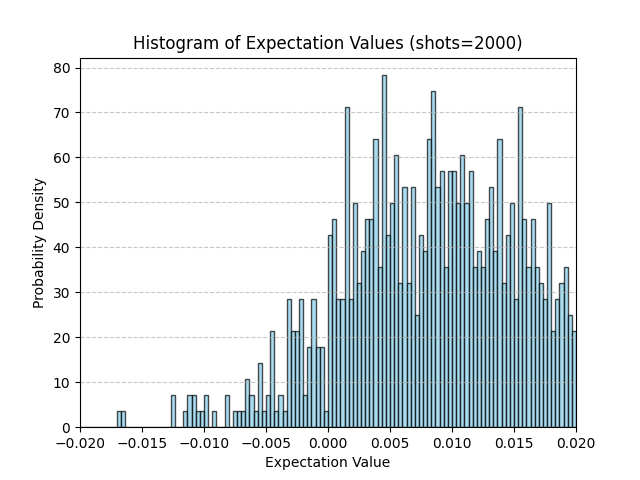}
		\caption{One saboteur}
		\label{fig:sub-integrity-1-raw}
	\end{subfigure}
	
	\caption{Integrity evaluation of an unprotected circuit cutting algorithm against a single attacker.}
    \Description{Integrity evaluation of an unprotected circuit cutting algorithm against a single attacker.}
	\label{fig:integrity-saboteurs-raw}
\end{figure}

\begin{figure}[htbp]
    \centering
    \begin{minipage}[c]{0.45\textwidth}
        \centering
        \begin{tabular}{|c|c|}
            \hline
            \textbf{Attackers} & \textbf{Distance (GT)} \\ [0.5ex]
            \hline
            0 & 0 \\ \hline
            1 & 0.785 \\ \hline
            2 & 0.849 \\ \hline
            3 & 0.857 \\ \hline
            4 & 0.833 \\ \hline
            5 & 0.913 \\ \hline
            6 & 0.788 \\ \hline
        \end{tabular}
        \caption{Hellinger distance from GT with growing attacking \ac{QPUs}.}
        \label{table:hellinger-raw-conf}
    \end{minipage}
    \hfill
    \begin{minipage}[c]{0.45\textwidth}
        \centering
        \includegraphics[width=\linewidth]{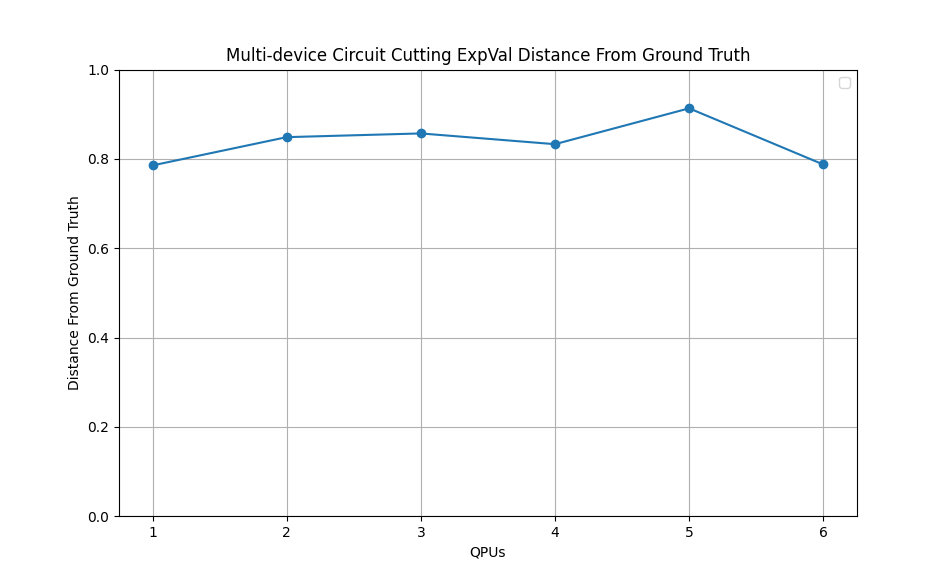}
        \caption{Hellinger distance with integrity saboteurs and unprotected cutting.}
        \label{fig:hellinger-integrity-raw}
    \end{minipage}
    \Description{Hellinger distance with integrity saboteurs and unprotected cutting.}
\end{figure}

As shown in the histograms in Figure~\ref{fig:integrity-saboteurs-raw} and by the variations on the Hellinger distance from the ground truth displayed in Figure~\ref{fig:hellinger-integrity-raw} and Table~\ref{table:hellinger-raw-conf}, even a single saboteur has a significant impact on the overall result, making it markedly different from the reference values. In other words, without the introduction of suitable countermeasures (see Section~\ref{subsec:countermeasures}) circuit cutting is not resilient against active attackers that compromise data integrity. Therefore, the subsequent tests will focus on enhancing integrity resilience through the application of several protective countermeasures.

\subsubsection{Second Test: Proportional Probability and 2X Replication}

In the second batch of tests, the sub-circuits, replicated twice, are distributed among the \ac{QPUs} with a probability proportional to their \textit{integrity score} $IS_{n}$ computed with probe circuits, as described in Section~\ref{subsec:countermeasures}. 
The probability $P_{n}$ of selecting the $n^{\text{th}}$ \ac{QPU} is computed as follows:

\begin{equation}
	\dfrac{IS_{n}}{\sum_{i=0}^{5} IS_{i}}.
\end{equation}

\begin{figure}[htbp]
	\centering
	\begin{subfigure}{0.45\textwidth}
		\centering
		\includegraphics[width=\textwidth]{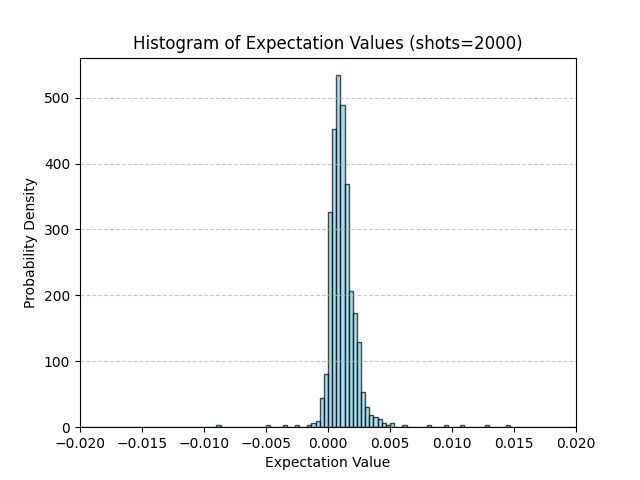}
		\caption{Three saboteurs}
		\label{fig:sub-integrity-3}
	\end{subfigure}
	\hfill
	\begin{subfigure}{0.45\textwidth}
		\centering
		\includegraphics[width=\textwidth]{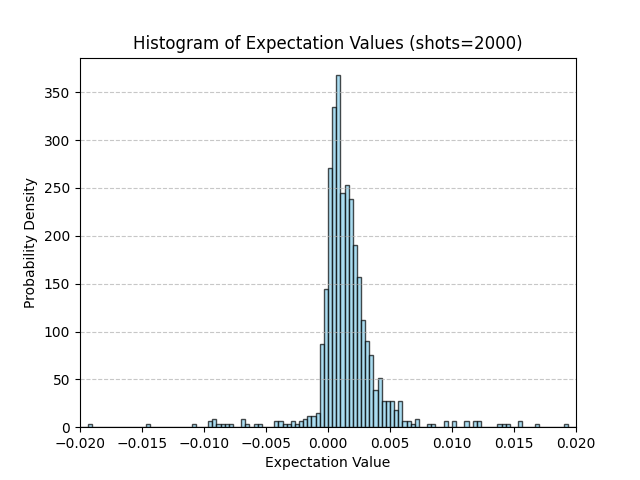}
		\caption{Four saboteurs}
		\label{fig:sub-integrity-4}
	\end{subfigure}
	
	\begin{subfigure}{0.45\textwidth}
		\centering
		\includegraphics[width=\textwidth]{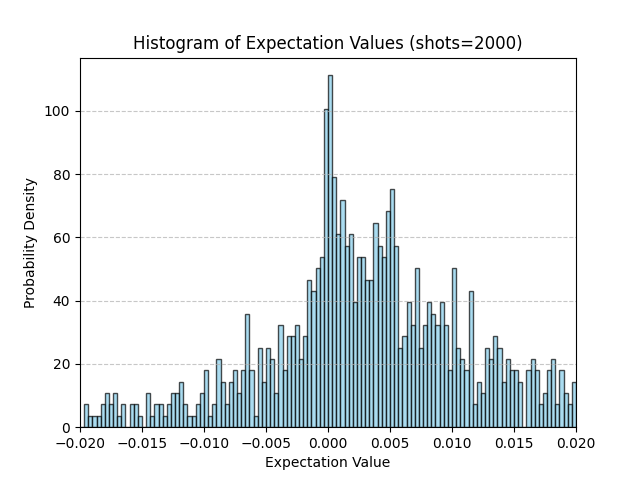}
		\caption{Five saboteurs}
		\label{fig:sub-integrity-5}
	\end{subfigure}
	\hfill
	\begin{subfigure}{0.45\textwidth}
		\centering
		\includegraphics[width=\textwidth]{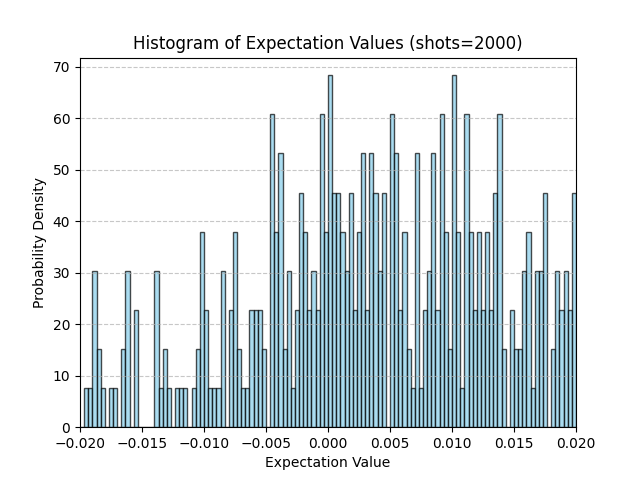}
		\caption{Six saboteurs}
		\label{fig:sub-integrity-6}
	\end{subfigure}
	
	\caption{Integrity evaluation with a progressively larger amount of saboteurs and with an execution probability proportional to the \ac{QPU} integrity score and a 2X replication factor.}
    \Description{Integrity evaluation with a progressively larger amount of saboteurs and with an execution probability proportional to the \ac{QPU} integrity score and a 2X replication factor.}
	\label{fig:integrity-saboteurs}
\end{figure}

Figure~\ref{fig:integrity-saboteurs} displays the high-level output probability distributions as the number of saboteurs increases (from 3 to 6). While a single saboteur does not significantly alter the overall distribution shape, the test involving six saboteurs results in a significantly different distribution.

\begin{figure}[htbp]
    \centering
    \begin{minipage}[c]{0.48\textwidth}
        \captionsetup{type=table}
        \centering
        \begin{tabular}{|c|c|}
            \hline
            \textbf{Attackers} & \textbf{Distance (GT)} \\ [0.5ex]
            \hline
            0 & 0 \\ \hline
            1 & 0.092 \\ \hline
            2 & 0.210 \\ \hline
            3 & 0.184 \\ \hline
            4 & 0.384 \\ \hline
            5 & 0.745 \\ \hline
            6 & 0.807 \\ \hline
        \end{tabular}
        \caption{Hellinger distance from GT with execution probability proportional to \ac{QPU} integrity and 2X replication.}
        \label{table:hellinger-integrity}
    \end{minipage}
    \begin{minipage}[c]{0.48\textwidth}
        \captionsetup{type=figure}
        \centering
        \includegraphics[width=\linewidth]{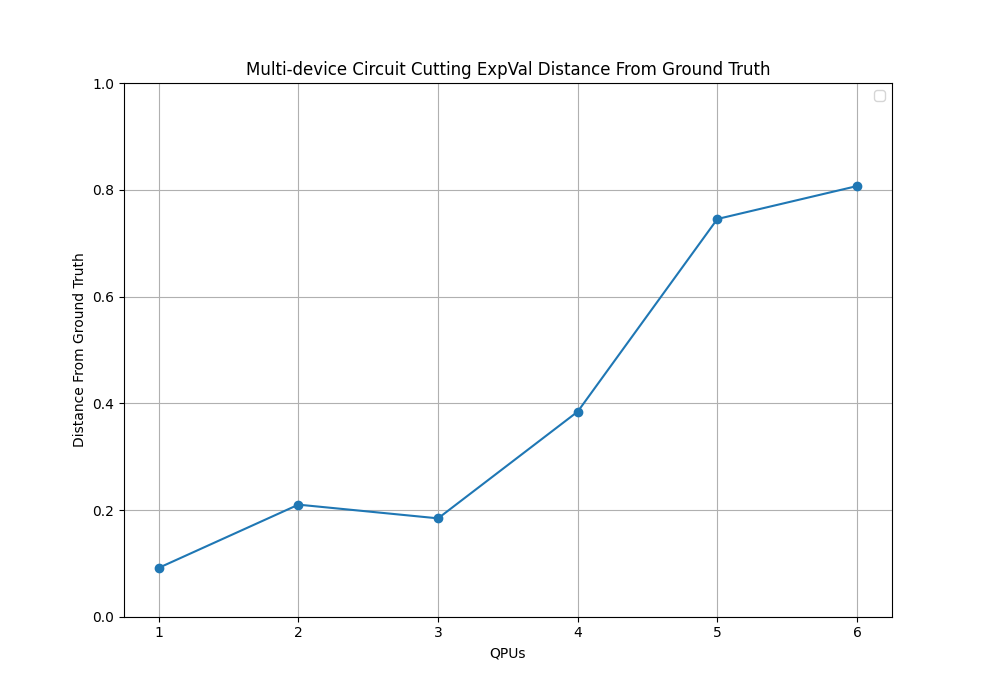}
        \caption{Hellinger distance from GT with integrity saboteurs, proportional execution probability, and 2X replication.}
        \label{fig:hellinger-integrity}
    \end{minipage}
    \Description{Hellinger distance from GT with integrity saboteurs, proportional execution probability, and 2X replication.}
\end{figure}

This visual result can be quantitatively assessed by measuring the Hellinger distance between each distribution and the ground truth. As shown in Figure~\ref{fig:hellinger-integrity} and in Table~\ref{table:hellinger-integrity}, the presence of up to three saboteurs among the six \ac{QPUs} does not significantly alter the outcome. However, from four to six saboteurs, the distance from the ground truth increases incrementally, approaching a value of 1.

\subsubsection{Third Test: Exponential Probability And 2X Replication}

In the third round of testing, the probability of assigning sub-circuits to the \ac{QPUs} was modified from being linearly proportional to being exponentially dependent on the \textit{integrity score} $IS_{n}$. Meanwhile, the replication factor was kept constant at 2X. The updated formula that determines which \ac{QPU} will be used to execute quantum sub-circuits is as follows:

\begin{equation}
	\dfrac{e^{IS_{n}}}{\sum_{i=0}^{5} e^{IS_{i}}}.
\end{equation}

\begin{figure}[htbp]
	\centering
	\begin{subfigure}{0.45\textwidth}
		\centering
		\includegraphics[width=\textwidth]{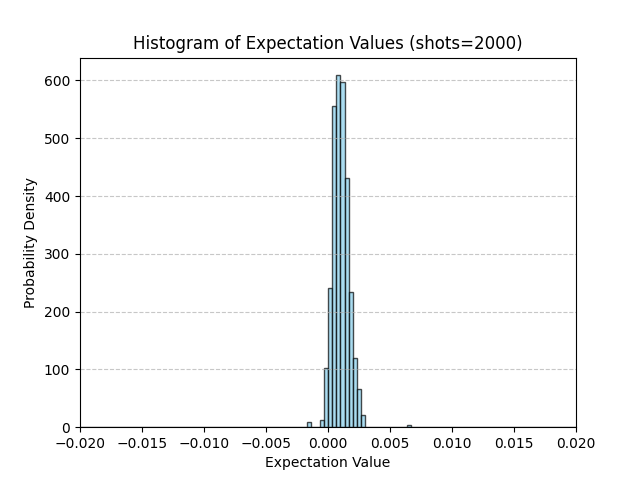}
		\caption{Three saboteurs}
		\label{fig:sub-integrity-3-exp}
	\end{subfigure}
	\hfill
	\begin{subfigure}{0.45\textwidth}
		\centering
		\includegraphics[width=\textwidth]{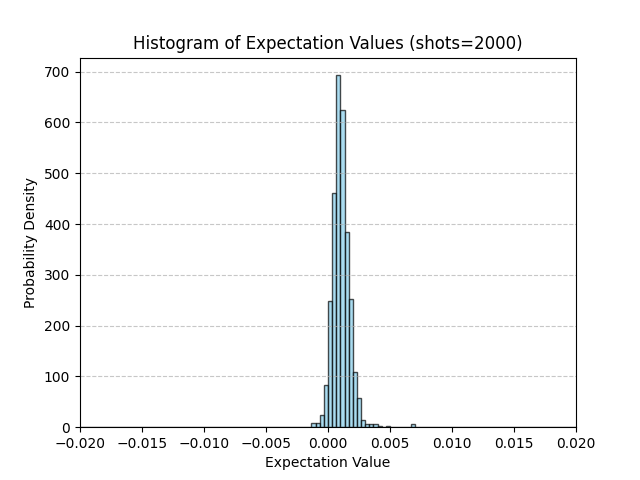}
		\caption{Four saboteurs}
		\label{fig:sub-integrity-4-exp}
	\end{subfigure}
	
	\begin{subfigure}{0.45\textwidth}
		\centering
		\includegraphics[width=\textwidth]{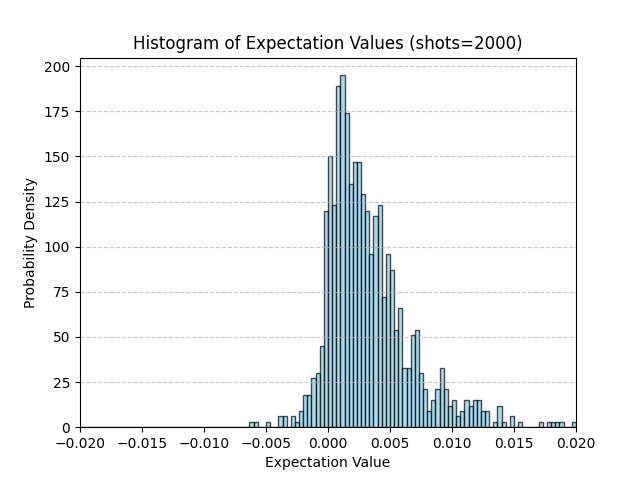}
		\caption{Five saboteurs}
		\label{fig:sub-integrity-5-exp}
	\end{subfigure}
	\hfill
	\begin{subfigure}{0.45\textwidth}
		\centering
		\includegraphics[width=\textwidth]{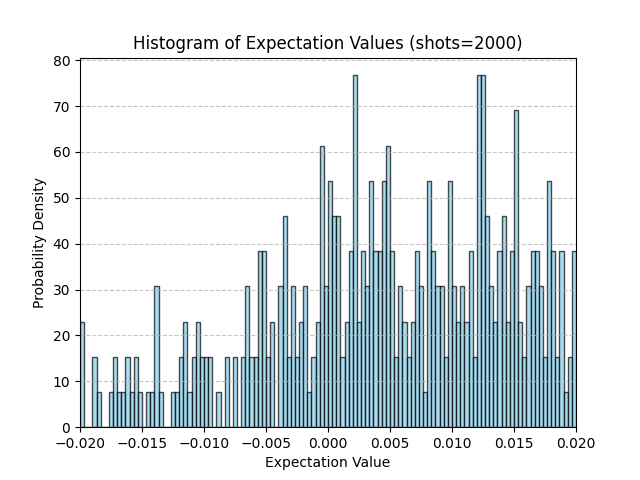}
		\caption{Six saboteurs}
		\label{fig:sub-integrity-6-exp}
	\end{subfigure}
	
	\caption{Integrity evaluation with a progressively larger amount of saboteurs with an execution probability exponential with respect to the \ac{QPU} integrity score and a 2X replication factor.}
    \Description{Integrity evaluation with a progressively larger amount of saboteurs with an execution probability exponential with respect to the QPU integrity score and a 2X replication factor.}
	\label{fig:integrity-saboteurs-exp}
\end{figure}

This assignment criterion shifts most of the workload to the most reliable \ac{QPUs} while under-utilising suspicious processors. Figure~\ref{fig:integrity-saboteurs-exp} displays the high-level output probability distributions with a growing number of saboteurs (from 3 to 6) and an exponential probability of sub-circuit execution.

\begin{figure}[htbp]
    \centering
    \begin{minipage}[c]{0.48\textwidth}
        \captionsetup{type=table}
        \centering
        \begin{tabular}{|c|c|}
            \hline
            \textbf{Attackers} & \textbf{Distance (GT)} \\ [0.5ex]
            \hline
            0 & 0 \\ \hline
            1 & 0.077 \\ \hline
            2 & 0.063 \\ \hline
            3 & 0.074 \\ \hline
            4 & 0.080 \\ \hline
            5 & 0.572 \\ \hline
            6 & 0.817 \\ \hline
        \end{tabular}
        \caption{Hellinger distance from GT with exponential execution probability and 2X replication.}
        \label{table:hellinger-integrity-exp-redundancy2}
    \end{minipage}
    \begin{minipage}[c]{0.48\textwidth}
        \captionsetup{type=figure}
        \centering
        \includegraphics[width=\linewidth]{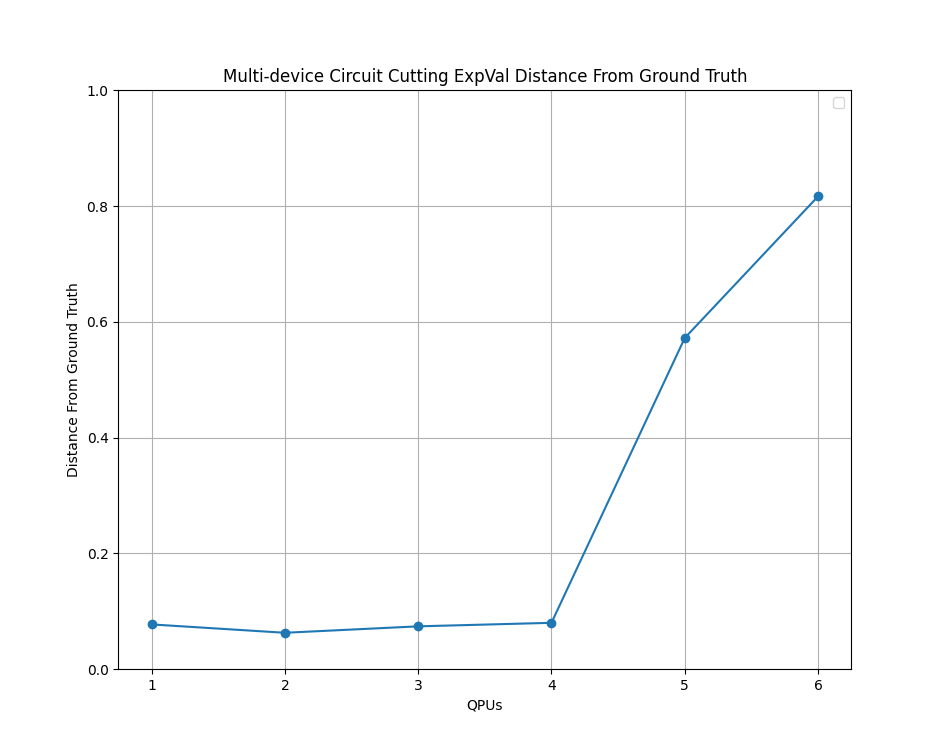}
        \caption{Hellinger distance from GT with integrity saboteurs, exponential execution probability, and 2X replication.}
        \label{fig:hellinger-integrity-exp}
        \Description{Hellinger distance from ground truth with a growing number of integrity saboteurs with an execution probability exponential with respect to the QPU integrity score and a 2X replication factor.}
    \end{minipage}
\end{figure}

This visual result can be quantitatively assessed by measuring the Hellinger distance between each distribution and the ground truth. As shown in Figure~\ref{fig:hellinger-integrity-exp} and Table~\ref{table:hellinger-integrity-exp-redundancy2}, the presence of up to four saboteurs among the six \ac{QPUs} (i.e., the majority of the \ac{QPUs} available) does not significantly alter the outcome. However, the distance increases sharply from five to six saboteurs, approaching a value of 1. This represents a clear improvement over the previous test, in which the assignment probability was linearly proportional to the \ac{QPU} integrity score.

\subsubsection{Fourth Test: Exponential Probability Without Replication}

Introducing exponential replication probabilities led to an improvement in integrity resilience; therefore, we will retain this probability criterion. We now proceed to evaluate the benchmark circuit \textit{without sub-circuit replication}. The probability model for sub-circuit distribution remains exponential with respect to the integrity score.

\begin{figure}[htbp]
    \centering
    \begin{minipage}[c]{0.48\textwidth}
        \captionsetup{type=table}
        \centering
        \begin{tabular}{|c|c|}
            \hline
            \textbf{Attackers} & \textbf{Distance (GT)} \\ [0.5ex]
            \hline
            0 & 0 \\ \hline
            1 & 0.080 \\ \hline
            2 & 0.068 \\ \hline
            3 & 0.077 \\ \hline
            4 & 0.098 \\ \hline
            5 & 0.114 \\ \hline
            6 & 0.752 \\ \hline
        \end{tabular}
        \caption{Hellinger distance from GT with exponential execution probability and no sub-circuit replication.}
        \label{table:hellinger-integrity-exp-redundancy1}
    \end{minipage}
    \hfill
    \begin{minipage}[c]{0.48\textwidth}
        \captionsetup{type=figure}
        \centering
        \includegraphics[width=\linewidth]{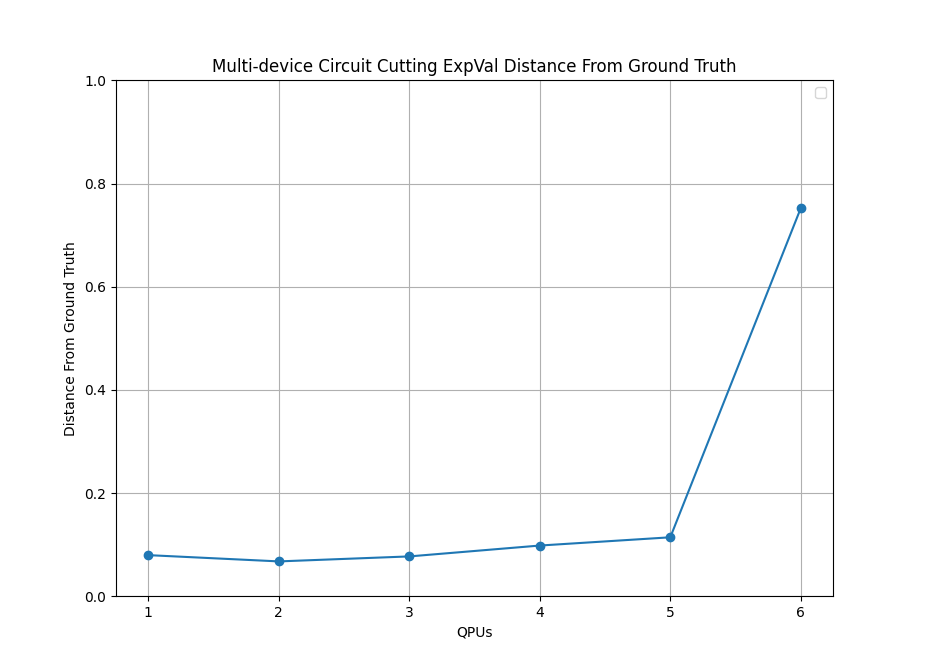}
        \caption{Hellinger distance from GT with integrity saboteurs, exponential execution probability, and no replication.}
        \label{fig:hellinger-integrity-exp-norep}
        \Description{Hellinger distance from ground truth with a growing number of integrity saboteurs with an execution probability exponential with respect to the QPU integrity score and without sub-circuit replication.}
    \end{minipage}
\end{figure}

As shown in Figure~\ref{fig:hellinger-integrity-exp-norep} and in Table~\ref{table:hellinger-integrity-exp-redundancy1}, this configuration yields the best possible result: the distance from the ground truth remains low with up to five saboteurs (i.e., even if more than half of our actors are malicious), with the only significantly altered outcome occurring when all six \ac{QPUs} are compromised.

\begin{figure}[htbp]
	\centering
	
	\begin{subfigure}{0.45\textwidth}
		\centering
		\includegraphics[width=\textwidth]{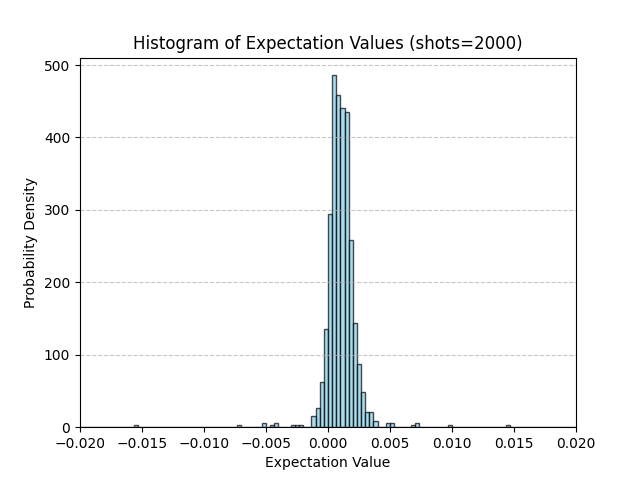}
		\caption{Five saboteurs}
        \Description{Five saboteurs}
		\label{fig:sub-integrity-5-exp-norep}
	\end{subfigure}
	\hfill
	\begin{subfigure}{0.45\textwidth}
		\centering
		\includegraphics[width=\textwidth]{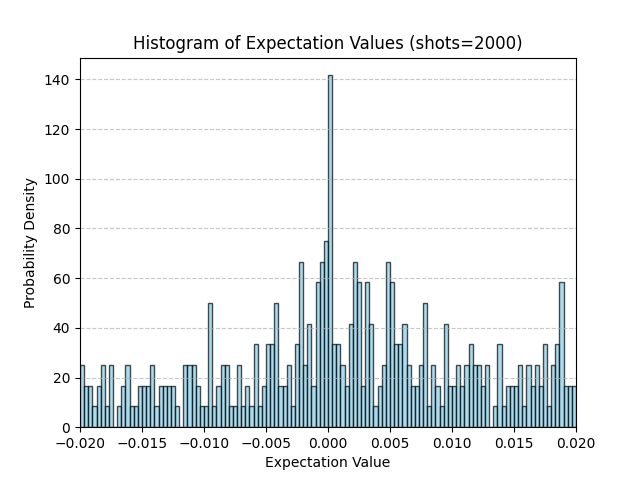}
		\caption{Six saboteurs}
        \Description{Six saboteurs}
		\label{fig:sub-integrity-6-exp-norep}
	\end{subfigure}
	
	\caption{Integrity evaluation with a progressively larger amount of saboteurs with an execution probability exponential with respect to the \ac{QPU} integrity score and without sub-circuit replication.}
    \Description{Integrity evaluation with a progressively larger amount of saboteurs with an execution probability exponential with respect to the \ac{QPU} integrity score and without sub-circuit replication.}
	\label{fig:integrity-saboteurs-exp-norep}
\end{figure}

\subsubsection{Further tests: Greenberger-Horne-Zeilinger Circuit}
\label{subsec:ghz}

The \ac{GHZ} circuit is used to create an entangled, highly-correlated state \cite{greenberger2007goingbellstheorem}. The \ac{GHZ} state is a generalisation of the Bell state for more than two qubits, and it is a cornerstone example in entanglement and non-local quantum phenomena. For example, in a three-qubit system, the \ac{GHZ} state is defined as follows:

\[
\ket{GHZ} = \dfrac{1}{\sqrt{2}}(\ket{000} + \ket{111}).
\]

The GHZ circuit generates the corresponding state with a combination of quantum gates. For example, in the case of three qubits, the circuit structure is the following:

\begin{enumerate}
	\item apply a Hadamard gate to the first qubit to create a $\ket{0} + \ket{1}$ quantum superposition: $H\ket{0} = \dfrac{1}{\sqrt{2}}(\ket{0} + \ket{1})$;
	\item apply two $CNOT$ gates, the first using the first qubit as control and the second qubit as target, the second using the first qubit as control and the third as target.
\end{enumerate}

Figure~\ref{fig:ghz3} depicts this circuit with a three qubit input.

\begin{figure}[htbp]
	\begin{center}
		\includegraphics[width=0.5\textwidth]{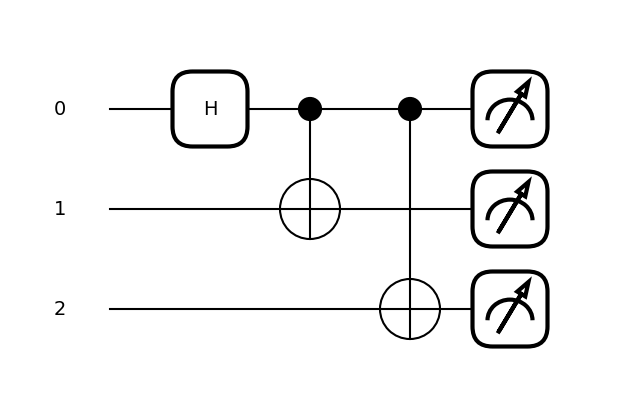}
		\caption{Three qubit \ac{GHZ} circuit.\label{fig:ghz3}}
        \Description{Three qubit \ac{GHZ} circuit.\label{fig:ghz3}}
	\end{center}
\end{figure}

The same tests conducted on the benchmark circuit were also performed on a 15-qubit \ac{GHZ} circuit from the \ac{MQT} benchmark suite \cite{quetschlich2023mqtbench} (with two circuit cuts applied):
\begin{enumerate}
	\item raw circuit cutting;
	\item proportional sub-circuit allocation probability and 2X replication;
	\item exponential sub-circuit allocation probability and 2X replication;
	\item exponential sub-circuit allocation probability and 3X replication;
	\item exponential sub-circuit allocation probability without replication.
\end{enumerate}

The tests were performed with a growing number of attacking \ac{QPUs}, ranging from 1 to 6, and the corresponding distribution of expectation values was compared to the ground truth (the circuit executed without any compromised \ac{QPU}) with the Hellinger Distance.

\begin{figure}[htbp]
	\centering
	\begin{subfigure}{0.45\textwidth}
		\centering
		\includegraphics[width=\textwidth]{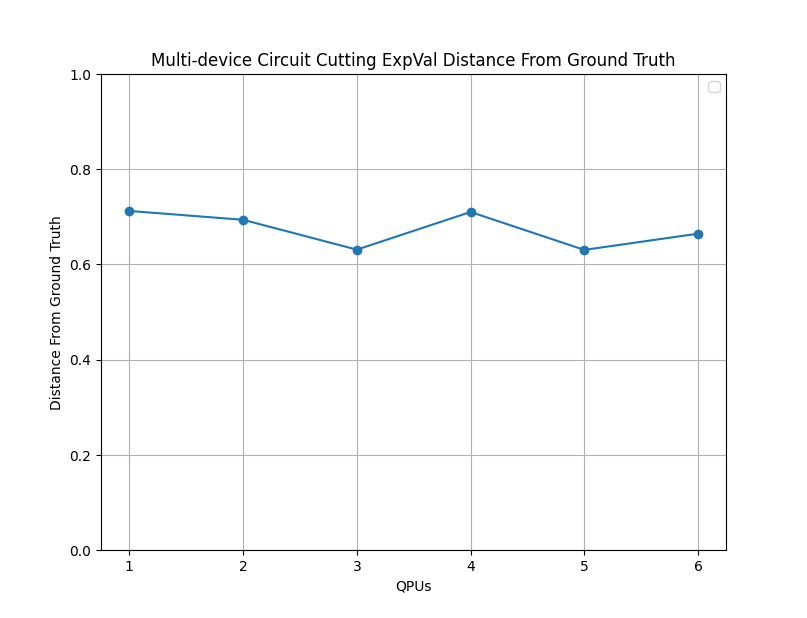}
		\caption{Raw circuit cutting test results}
		\label{fig:sub-integrity-ghz-raw}
	\end{subfigure}
	\hfill
	\begin{subfigure}{0.45\textwidth}
		\centering
		\includegraphics[width=\textwidth]{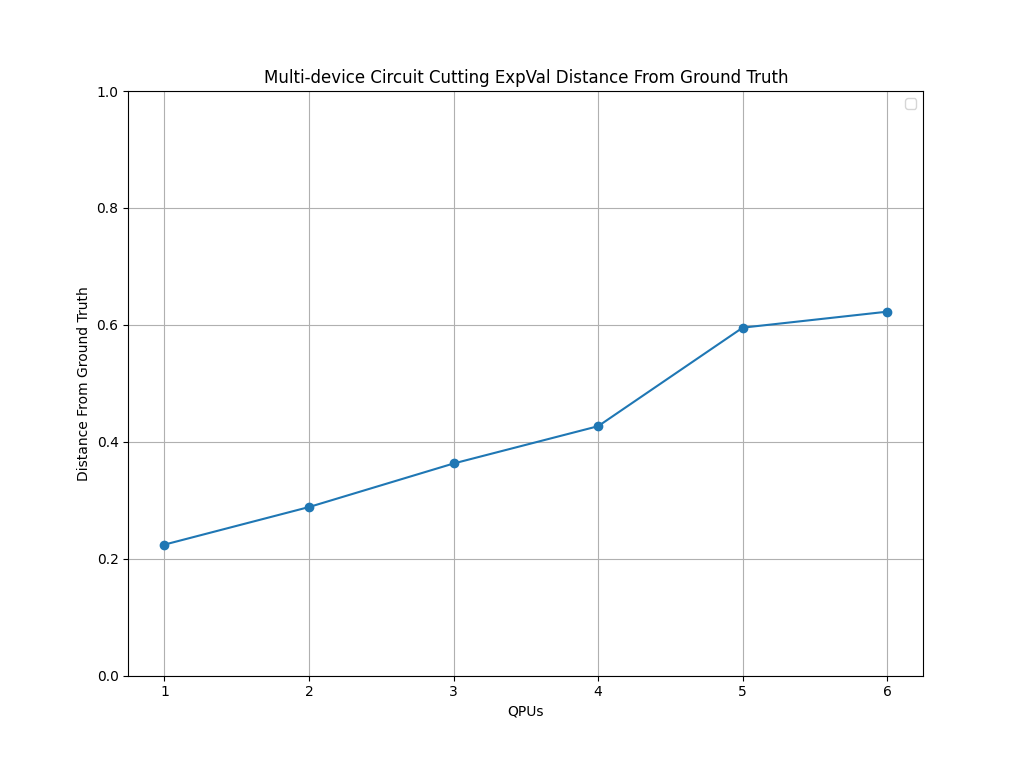}
		\caption{Proportional probability and 2X replication test results}
		\label{fig:sub-integrity-ghz-prop2x}
	\end{subfigure}
	\begin{subfigure}{0.45\textwidth}
		\centering
		\includegraphics[width=\textwidth]{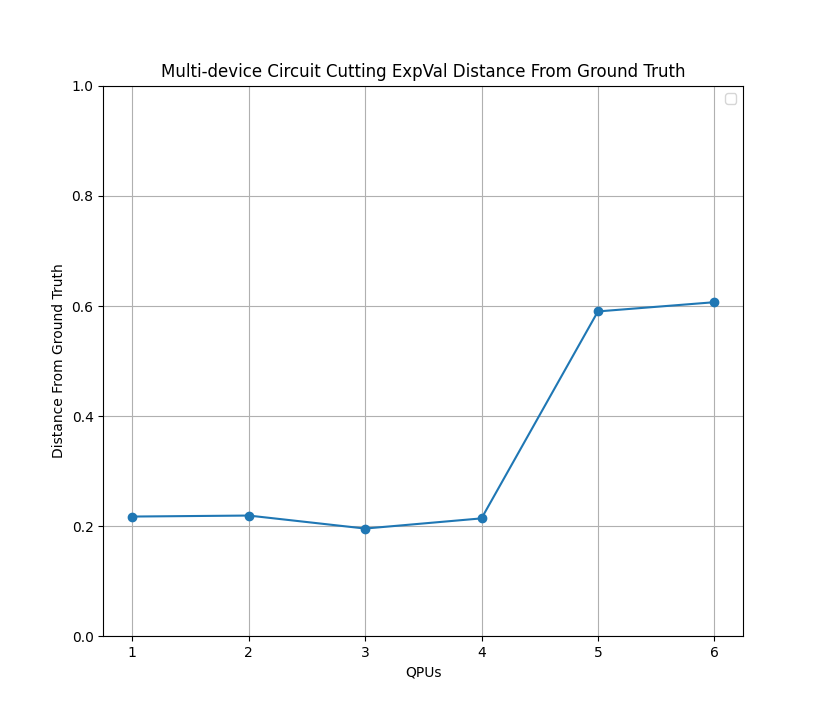}
		\caption{Exponential probability and 2X replication test results}
		\label{fig:sub-integrity-ghz-exp2x}
	\end{subfigure}
	\hfill
	\begin{subfigure}{0.45\textwidth}
		\centering
		\includegraphics[width=\textwidth]{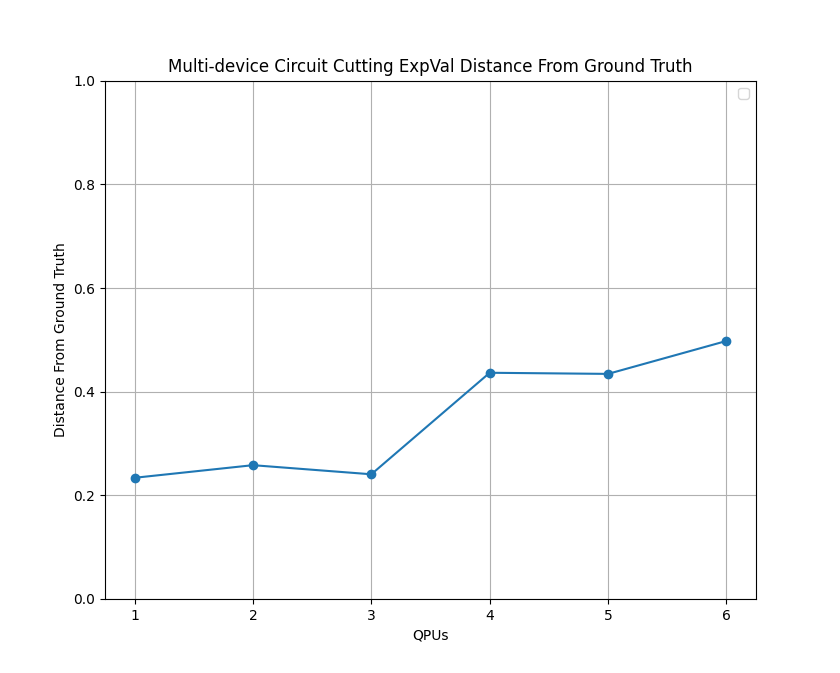}
		\caption{Exponential probability and 3X replication test results}
		\label{fig:sub-integrity-ghz-exp3x}
	\end{subfigure}
	
	\begin{subfigure}{0.45\textwidth}
		\centering
		\includegraphics[width=\textwidth]{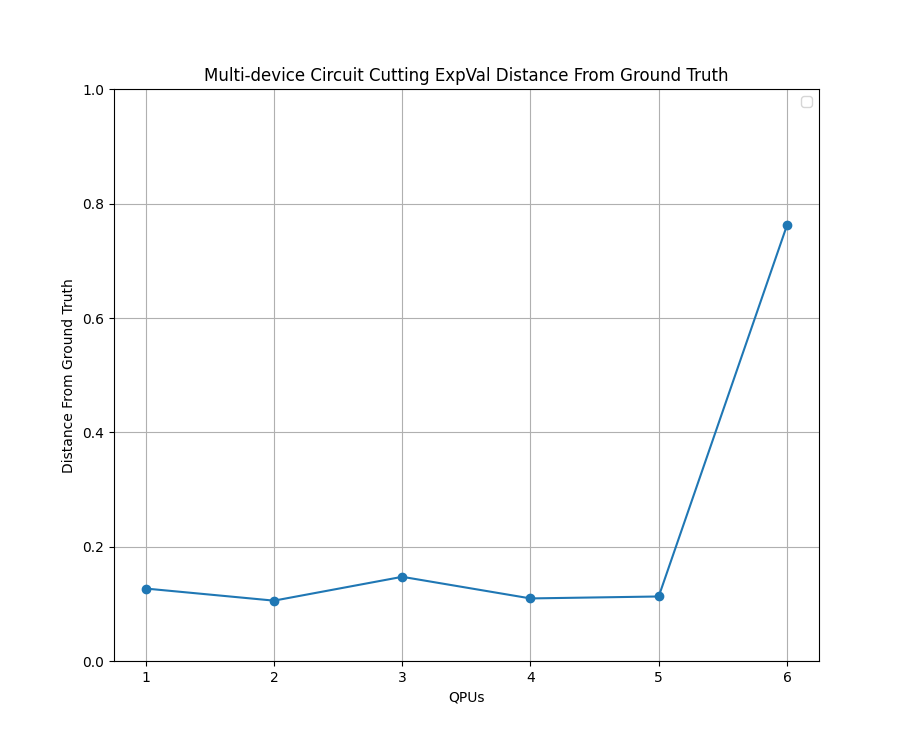}
		\caption{Exponential probability without replication test results}
		\label{fig:sub-integrity-ghz-exp1x}
	\end{subfigure}
	
	\caption{Integrity resilience tests on the \ac{GHZ} circuit (Hellinger distance from ground truth with a growing number of attackers).}
    \Description{Integrity resilience tests on the \ac{GHZ} circuit (Hellinger distance from ground truth with a growing number of attackers).}
	\label{fig:hellinger-integrity-ghz}
\end{figure}

Figure~\ref{fig:hellinger-integrity-ghz} shows the test results: the trend, summarised in Table~\ref{table:integrity-ghz-recap}, show the same behaviour as the benchmark circuit tests.

\begin{table}[ht]
	\centering
	\begin{tabular}{|l|c|}
		\hline
		\textbf{Configuration} & \textbf{Tolerated Attackers (0-6)} \\ [0.5ex]
		\hline
		Raw circuit cutting & 0 \\ \hline
		Proportional probability and 2X replication & 1 \\ \hline
		Exponential probability and 2X replication & 4 \\ \hline
		Exponential probability and 3X replication & 3 \\ \hline
		Exponential probability without replication & 5 \\ \hline
	\end{tabular}
	\caption{Summary of integrity-related experimental results on the \ac{GHZ} circuit.}
	\label{table:integrity-ghz-recap}
\end{table}

\subsubsection{Further tests: Deutsch-Jozsa Algorithm Circuit}

The Deutsch-Jozsa algorithm solves the following problem: given an oracle that implements a function $f : {0,1}^{n} \rightarrow {0,1}$, determine whether $f$ is \textit{constant}: if $f$ is constant, it will return the same value -- either $0$ or $1$ -- no matter the input; if $f$ is \textit{balanced} it will return $0$ for exactly half of the possible input and $1$ for the other half.

The standard deterministic algorithm would require, in the worst case, at least $2^{(n-1)} + 1$ evaluations of $f$, where $n$ is the number of bits. The quantum Deutsch-Jozsa algorithm, instead, provides the solution with a single evaluation of $f$.

The quantum algorithm works as follows:

\begin{enumerate}
	\item Define the initial state by using a $n$ qubit register initialised to $\ket{0}$ and an auxiliary qubit initialised to $\ket{1}$:
	\[ 
	\ket{0}^{\otimes n} \otimes \ket{1}.
	\]
	\item Apply a Hadamard gate $H$ to each of the $n + 1$ qubits to create a quantum superposition in which all possible inputs are represented:
	\[
	\frac{1}{\sqrt{2^n}} \sum_{x=0}^{2^n-1} \ket{x} \otimes \frac{\ket{0} - \ket{1}}{\sqrt{2}}.
	\]
	\item Call the quantum oracle, implemented as a $U_f$ port that acts on the qubits with the following transformation:
	 \[
	 U_f: \ket{x}\ket{y} \to \ket{x}\ket{y \oplus f(x)}.
	 \]
	\item After the oracle call, the information about $f(x)$ is encoded in the input registry phase as follows:
	\[
	\frac{1}{\sqrt{2^n}} \sum_{x=0}^{2^n-1} (-1)^{f(x)} \ket{x} \otimes \frac{\ket{0} - \ket{1}}{\sqrt{2}}.
	\]
	\item Apply again the Hadamard gate to the first $n$ qubits to transform the phases in amplitudes.
	\item Measure the final state of the first $n$ qubits: the resulting observation will always be $\ket{0}^{\otimes n}$ if the function is \textit{constant}, otherwise it is \textit{balanced}.
\end{enumerate}

\begin{figure}[htbp]
	\begin{center}
		\includegraphics[width=0.5\textwidth]{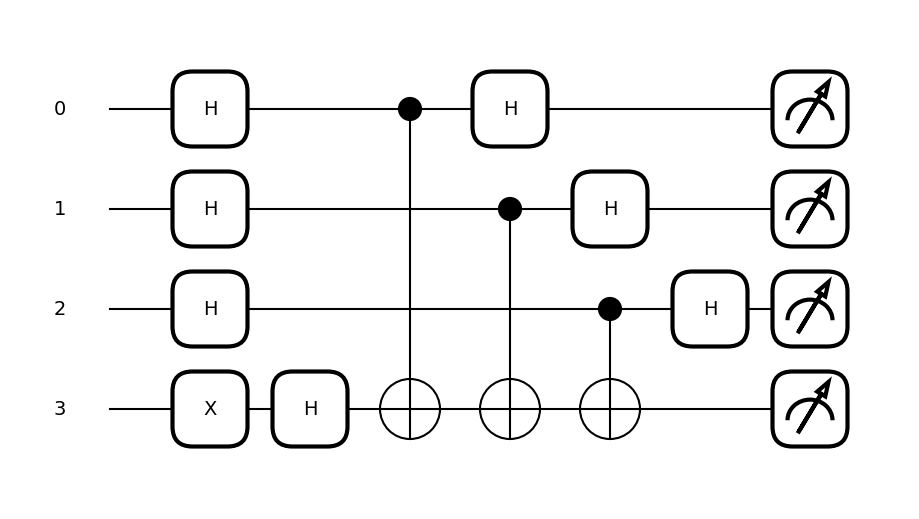}
		\caption{Deutsch-Jozsa circuit for a 3 bit function.\label{fig:dj3}}
        \Description{Deutsch-Jozsa circuit for a 3 bit function.}
	\end{center}
\end{figure}

Figure~\ref{fig:dj3} displays a Deutsch-Jozsa quantum circuit for a 3-bit function. It uses four qubits, including the auxiliary one.

The same tests performed on the benchmark circuit and the \ac{GHZ} circuit (see Subsection~\ref{subsec:ghz}) have also been applied to a 15 qubit Deutsch-Jozsa circuit from the \ac{MQT} benchmark suite \cite{quetschlich2023mqtbench} (with two circuit cuts applied):
\begin{enumerate}
	\item raw circuit cutting;
	\item proportional sub-circuit allocation probability and 2X replication;
	\item exponential sub-circuit allocation probability and 2X replication;
	\item exponential sub-circuit allocation probability and 3X replication;
	\item exponential sub-circuit allocation probability without replication.
\end{enumerate}

All tests have been executed with a growing number of attacking \ac{QPUs}, from 1 to 6, and the corresponding distribution of expectation values was compared with the ground truth (the circuit executed without any offending \ac{QPU}).

\begin{figure}[htbp]
	\centering
	\begin{subfigure}{0.45\textwidth}
		\centering
		\includegraphics[width=\textwidth]{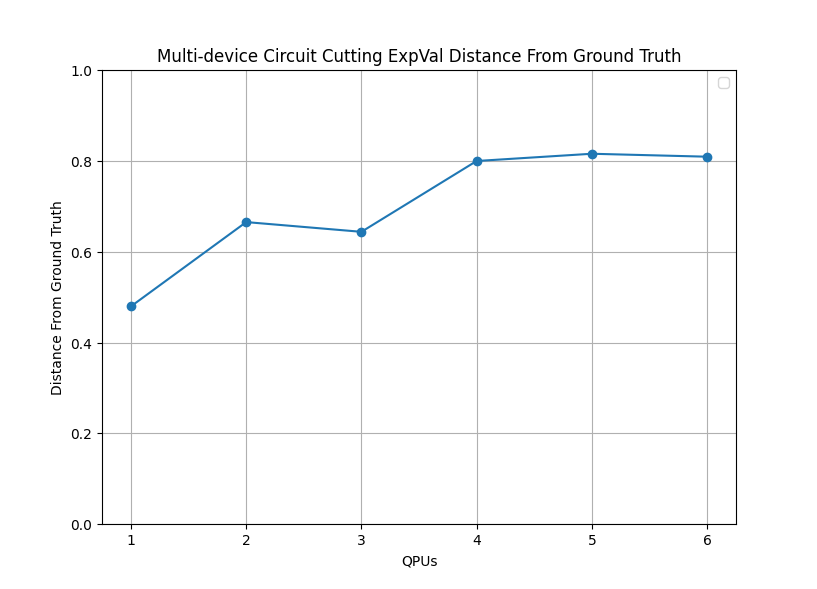}
		\caption{Raw circuit cutting test results}
		\label{fig:sub-integrity-dj-raw}
	\end{subfigure}
	\hfill
	\begin{subfigure}{0.45\textwidth}
		\centering
		\includegraphics[width=\textwidth]{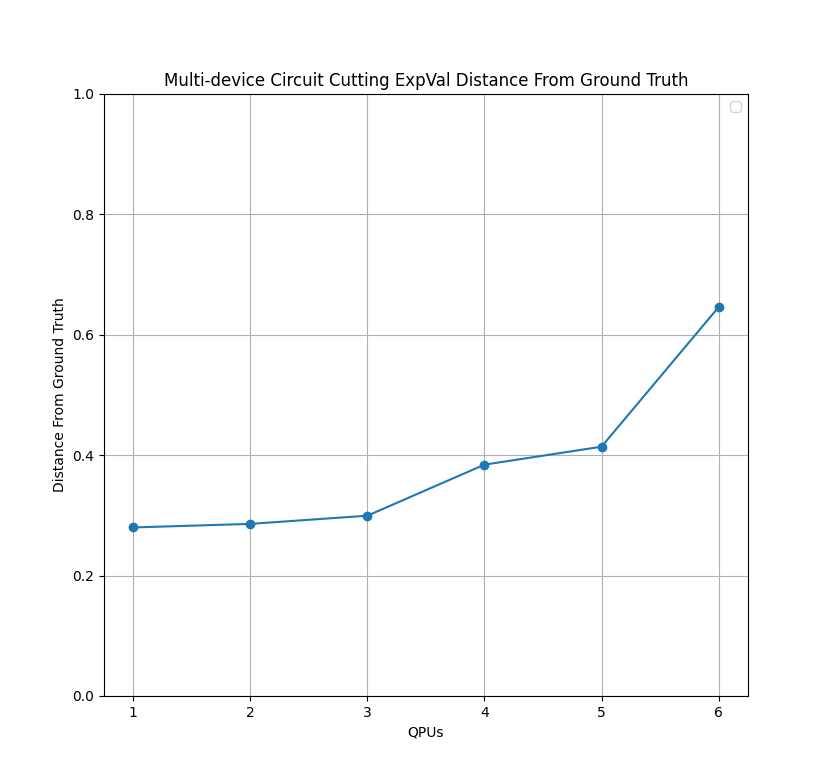}
		\caption{Proportional probability and 2X replication test results}
		\label{fig:sub-integrity-dj-prop2x}
	\end{subfigure}
	\begin{subfigure}{0.45\textwidth}
		\centering
		\includegraphics[width=\textwidth]{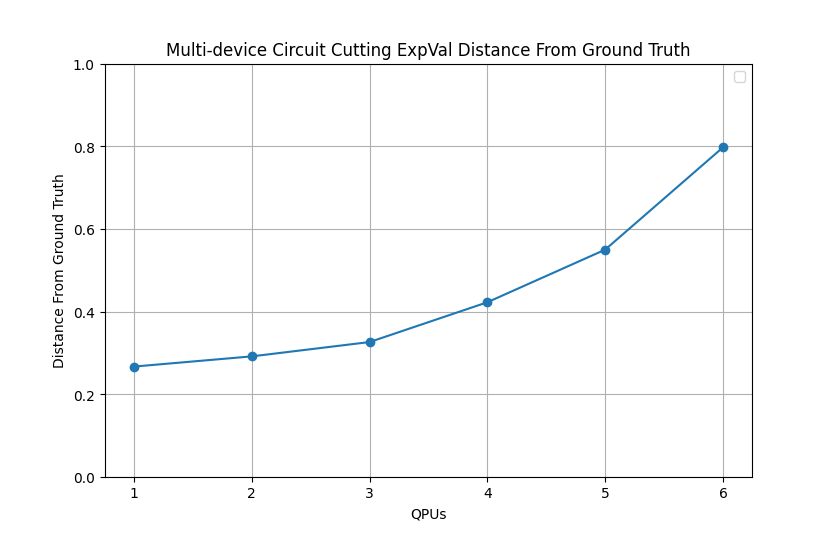}
		\caption{Exponential probability and 2X replication test results}
		\label{fig:sub-integrity-dj-exp2x}
	\end{subfigure}
	\hfill
	\begin{subfigure}{0.45\textwidth}
		\centering
		\includegraphics[width=\textwidth]{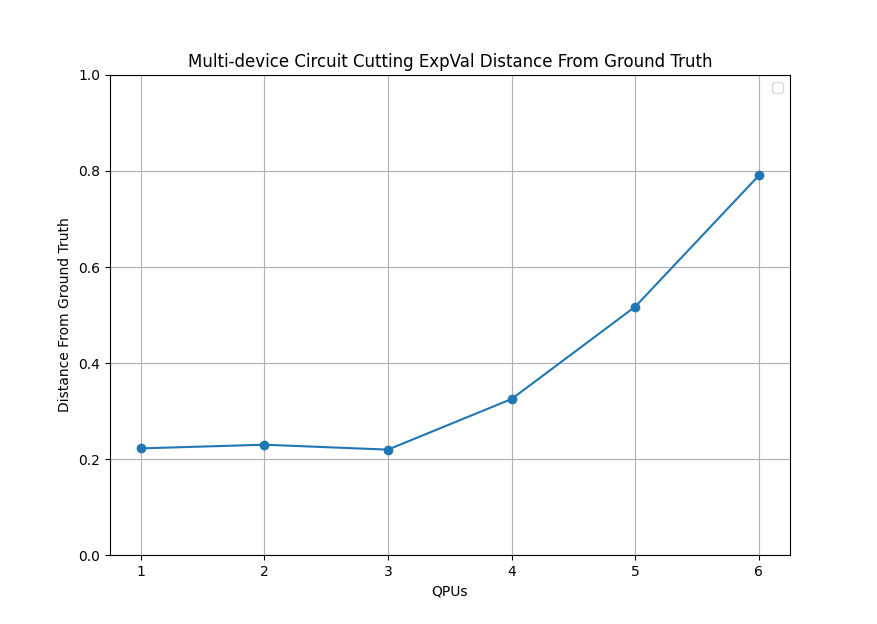}
		\caption{Exponential probability and 3X replication test results}
		\label{fig:sub-integrity-dj-exp3x}
	\end{subfigure}
	
	\begin{subfigure}{0.45\textwidth}
		\centering
		\includegraphics[width=\textwidth]{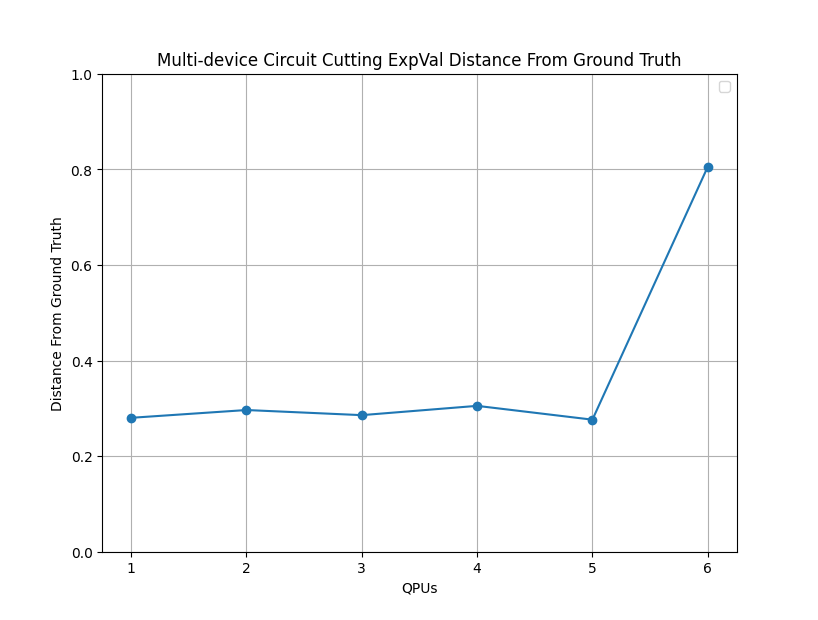}
		\caption{Exponential probability without replication test results}
        \Description{Exponential probability without replication test results}
		\label{fig:sub-integrity-dj-exp1x}
	\end{subfigure}
	
	\caption{Integrity resilience tests on the Deutsch-Jozsa circuit (Hellinger distance from ground truth with a growing number of attackers)}
	\label{fig:hellinger-integrity-dj}
\end{figure}

Figure~\ref{fig:hellinger-integrity-dj} shows the test results: the trend, summarised in Table~\ref{table:integrity-dj-recap}, is the same as that of the benchmark circuit tests.

\begin{table}[ht]
	\centering
	\begin{tabular}{|l|c|}
		\hline
		\textbf{Configuration} & \textbf{Tolerated Attackers (0-6)} \\ [0.5ex]
		\hline
		Raw circuit cutting & 0 \\ \hline
		Proportional probability and 2X replication & 2 \\ \hline
		Exponential probability and 2X replication & 2 \\ \hline
		Exponential probability and 3X replication & 3 \\ \hline
		Exponential probability without replication & 5 \\ \hline
	\end{tabular}
	\caption{Summary of integrity-related experimental results on the Deutsch-Jozsa circuit.}
	\label{table:integrity-dj-recap}
\end{table}

\subsubsection{Summary of the Results}
\label{subsec:discussion-integrity}

Table~\ref{table:integrity-recap} summarises the integrity-related experiments presented in this Section.

\begin{table}[ht]
	\centering
	\begin{tabular}{|l|c|c|c|}
		\hline
		\textbf{Configuration} & \multicolumn{3}{c|}{\textbf{Tolerated Attackers (0-6)}} \\ \hline
		& \textbf{Benchmark} & \textbf{GHZ} & \textbf{Deutsch-Jozsa} \\ [0.5ex]
		\hline
		Raw circuit cutting & 0 & 0 & 0\\ \hline
		Proportional probability and 2X replication & 3 & 1 & 2\\ \hline
		Exponential probability and 2X replication & 4 & 4 & 2\\ \hline
		Exponential probability and 3X replication & 3 & 3 & 3\\ \hline
		Exponential probability without replication & 5 & 5 & 5\\ \hline
	\end{tabular}
	\caption{Summary of integrity-related experimental results.}
	\label{table:integrity-recap}
\end{table}

It is therefore evident that \textbf{Goal G1} (stated in Subsection~\ref{subsec:objectives}), can be achieved through quantum circuit cutting, enhanced with sub-circuit allocation probabilities that are exponential with respect to dynamic integrity scores computed using suitable probe circuits while avoiding sub-circuit replication.

\subsubsection{Proportional VS Exponential Sub-Circuit Allocation Probability}
\label{subsubsec:probs}

Among the countermeasures described in Section~\ref{subsec:countermeasures}, the sub-circuit \ac{QPU} allocation probability emerges as the most critical factor in determining the optimal quantum circuit cutting configuration for integrity resilience. 

The tests in Section~\ref{subsec:integrity} used two different functions to determine the probability $P_{n}$ of selecting  $n^{\text{th}}$ \ac{QPU}:

\begin{itemize}
	\item \textbf{Proportional}: $\dfrac{IS_{n}}{\sum_{i=0}^{5} IS_{i}}$
	\item \textbf{Exponential}: $\dfrac{e^{IS_{n}}}{\sum_{i=0}^{5} e^{IS_{i}}}$
\end{itemize}

Both formulae take into account the \ac{QPUs} \textit{integrity scores} $IS$.

\begin{figure}[htbp]
	\begin{center}
		\includegraphics[height=\textheight]{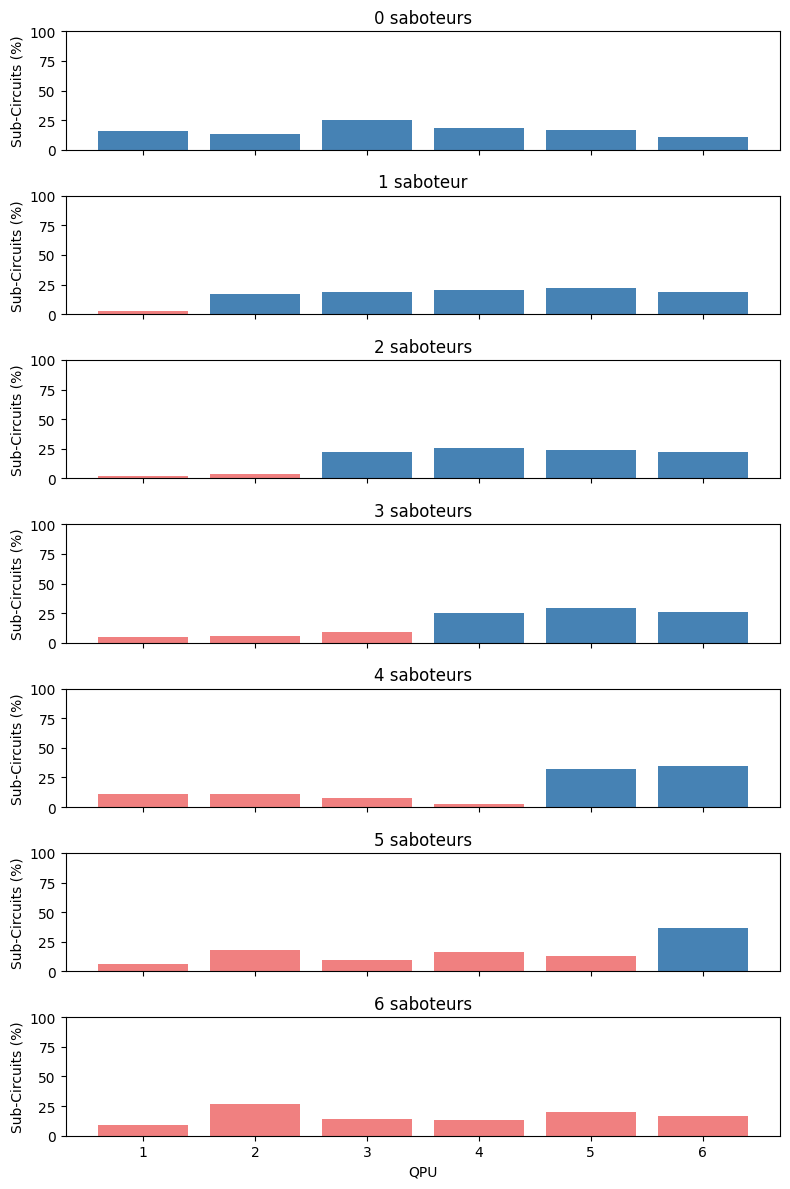}
		\caption{Proportional sub-circuit allocation probability with a growing number of saboteurs. The percentages have been computed by averaging the sub-circuit allocation rate of each experiment with proportional sub-circuit allocation probability, grouped by the number of active saboteurs.\label{fig:qpu-assignments-prop}}
        \Description{Proportional sub-circuit allocation probability with a growing number of saboteurs. The percentages have been computed by averaging the sub-circuit allocation rate of each experiment with proportional sub-circuit allocation probability, grouped by the number of active saboteurs.}
	\end{center}
\end{figure}

\begin{figure}[htbp]
	\begin{center}
		\includegraphics[height=\textheight]{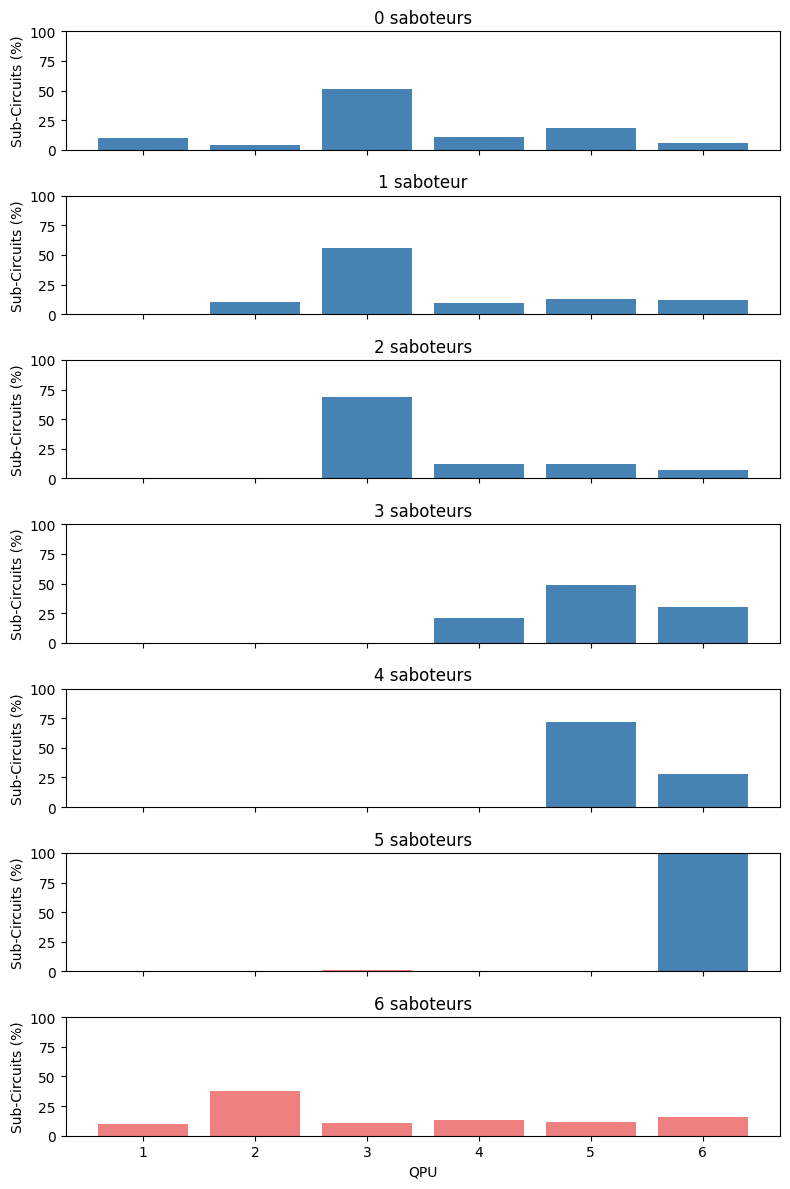}
		\caption{Exponential sub-circuit allocation probability with a growing number of saboteurs. The percentages have been computed by averaging the sub-circuit allocation rate of each experiment with exponential sub-circuit allocation probability, grouped by the number of active saboteurs.\label{fig:qpu-assignments-exp}}
        \Description{Exponential sub-circuit allocation probability with a growing number of saboteurs. The percentages have been computed by averaging the sub-circuit allocation rate of each experiment with exponential sub-circuit allocation probability, grouped by the number of active saboteurs.}
	\end{center}
\end{figure}

As discussed in subsection \ref{subsec:discussion-integrity}, the advantage of the exponential probability model is evident. To understand why, we examine the average \ac{QPU} sub-circuit allocation percentages under an increasing number of saboteurs, from 0 to 6. Figure~\ref{fig:qpu-assignments-prop} shows the allocation percentages for the proportional model, while Figure~\ref{fig:qpu-assignments-exp} shows those for the exponential model. The percentages were computed by averaging the sub-circuit allocation rate of each experiment, grouped by allocation policy (proportional or exponential) and number of active saboteurs.

Clearly, the exponential probability formula is much more aggressive than the proportional one, as it tends to underutilise even good-behaving \ac{QPUs} when their integrity score is just slightly lower than the best-performing ones. However, \ac{QPUs} with a very low score, and therefore defective or malicious, are guaranteed to have little to no sub-circuits assigned up to 5 attacking devices over 6. While the proportional probability formula ensures a more balanced allocation across all devices, it also results in poorly performing \ac{QPUs} consistently receiving a significant share of sub-circuits.

Overall, the rationale behind the choice of these models can be understood as a trade-off between integrity resilience (a cybersecurity preference) and load distribution (a performance and possibly cost preference). To fulfil \textbf{Goal G1} (ensuring integrity resilience), the exponential probability model appears to be the most effective choice; however, this may not always hold true in different contexts.

\subsection{Confidentiality Testing}
\label{subsec:confidentiality}

To fulfil \textbf{Goal G2} (see section \ref{subsec:objectives}), our experiments aim to verify whether circuit cutting (possibly enhanced with additional countermeasures) can ensure the confidentiality of the input circuit. Intuitively, circuit cutting alone can help invalidate parts of the circuit and output data from being exposed to untrusted \ac{QPUs}, simply by avoiding the allocation of all sub-circuits to a single \ac{QPU}.

However, our \ac{PNI}-inspired testing methodology requirement is more stringent: ideally, the \ac{QPUs} should not be able to distinguish between different inputs. In other words, they should not be able to pinpoint the actual computation that they have been tasked to perform. The experimental results summarised in this section, therefore, aim to measure the capability of malicious \ac{QPUs} to distinguish between different input circuits.

Note that the output data being analysed will be the complete set of individual \ac{QPUs} output values, possibly including fake circuit outputs, without reconstructing them through the post-processing algorithm. To understand the rationale behind it, we shall revisit the \ac{PNI} definition of confidentiality:

\begin{equation} \label{eq:pni-confidentiality2}
	P(O_{lo} | I_{hi}) \simeq P(O_{lo} | I_{hi'}),
\end{equation}

where:

\begin{itemize}
	\item $I_{hi}$ and $I_{hi'}$ are two high-level inputs fully available to the scheduler that malicious \ac{QPUs} are trying to distinguish;
	\item $O_{lo}$ is the low-level output produced by each \ac{QPU}.
\end{itemize}

The high-level output $O_{hi}$ obtained by combining the low-level outputs of the \ac{QPUs}, is not involved in the confidentiality testing.

Verifying whether Equation \ref{eq:pni-confidentiality2} holds is non-trivial for several reasons:

With a raw circuit cutting scheme, different high-level inputs $I_{hi}$ give rise to different sub-circuits, which in turn produce different low-level outputs $O_{lo}$. One possible way to mitigate this issue is to flood the \ac{QPUs} with fake sub-circuits, thereby diluting the actual results (see the corresponding additional countermeasure defined in Section~\ref{subsec:countermeasures}). However, demonstrating the effectiveness of this dummy sub-circuit flooding technique using quantitative techniques, such as histograms and Hellinger distances from the ground truth, is challenging. The fake values would need to fall within the expected range and approximate magnitude of legitimate outputs; otherwise, they risk being discarded or simply ignored. Note that, in the general case, it is not possible to know the correct output of the original quantum circuit evaluation prior to execution, which complicates the design of effective strategies.

To rigorously assess the effectiveness of the proposed confidentiality countermeasures, we structured the experimental campaign into three distinct phases:

\begin{enumerate}
    \item \textbf{Baseline Assessment:} We first verify whether two different input circuits ($I_{hi}$ and $I_{hi'}$) result in distinguishable low-level output distributions ($O_{lo}$ and $O_{lo'}$), establishing the vulnerability of raw circuit cutting.
    \item \textbf{Random Noise Injection:} We then evaluate the obfuscation provided by \textit{genuinely random} fake circuits, testing if unstructured noise is sufficient to hide the input.
    \item \textbf{Calibrated Noise Injection:} Finally, we assess the impact of \textit{artificially constructed (calibrated)} fake circuits, designed to produce results within the same magnitude as the real ones, to determine if targeted noise is required for true confidentiality.
\end{enumerate}

The number of fake circuits used in the tests will be determined by a multiplier $m$: for each legitimate sub-circuit, each \ac{QPU} will also have to process $m$ fake circuits (specifically, we test $m=5$ and $m=10$).

In confidentiality testing, attackers play a passive role and are typically undetectable. The scheduler user can provide \textit{confidentiality scores} based on information about \ac{QPUs} and their operators (see Section~\ref{subsec:countermeasures}). However, in the following tests, we will assume that all processors have equal confidentiality score.

\subsubsection{First Test: Raw Circuit Cutting}

It is not surprising that raw circuit cutting might provide the user with a certain degree of confidentiality, given that individual \ac{QPUs} receive only fragments of the original quantum circuit. However, the requirement that each \ac{QPU} be unable to distinguish between different high-level input circuits is, intuitively, not satisfied, as different input circuits inevitably result in different sub-circuits and, consequently, different output values.

\begin{figure}[htbp]
    \centering
    \begin{minipage}[c]{0.48\textwidth}
        \captionsetup{type=table}
        \centering
        \begin{tabular}{|l|c|}
            \hline
            \textbf{Circuit} & \textbf{Hellinger Dist.} \\ [0.5ex]
            \hline
            Alt. benchmark & 0.991 \\ \hline
            \ac{GHZ} & 0.923 \\ \hline
            Deutsch-Jozsa & 0.837 \\ \hline
        \end{tabular}
        \caption{Low-level output Hellinger distance from reference benchmark with raw circuit cutting.}
        \label{table:hellinger-raw}
    \end{minipage}
    \hfill
    \begin{minipage}[c]{0.48\textwidth}
        \captionsetup{type=figure}
        \centering
        \includegraphics[width=\linewidth]{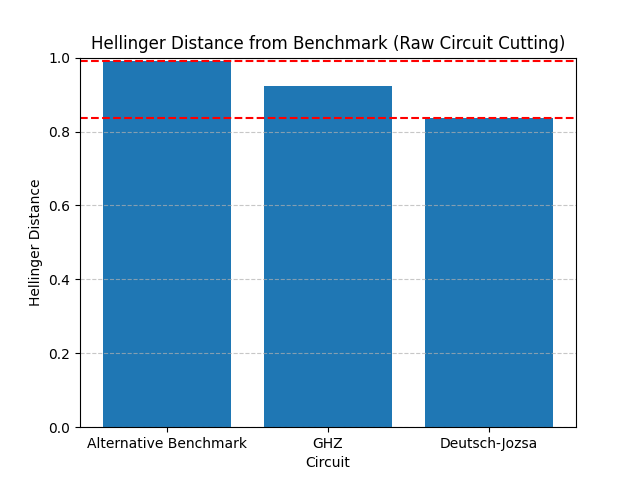}
        \caption{Hellinger distance from benchmark low-level \ac{QPU} output with raw circuit cutting (red lines delimit min/max range).}
        \label{fig:conf_raw_chart}
        \Description{Hellinger distance from benchmark circuit low-level \ac{QPU} output data with raw circuit cutting (lower is better).}
    \end{minipage}
\end{figure}

The numerical comparison of the Hellinger distances between the low-level \ac{QPUs} outputs of the benchmark circuits confirms this hypothesis. Table~\ref{table:hellinger-raw} and Figure~\ref{fig:conf_raw_chart} illustrate such distances, which are very close to the upper limit, meaning that a malicious \ac{QPU} can easily distinguish between different input circuits. Note that the \ac{PNI}-inspired constraint defined in Section~\ref{sec:approach_overview} is not satisfied.

\subsubsection{Second Test: 5X Fake Circuits}

\begin{figure}[htbp]
  \centering
  \begin{subfigure}[b]{0.48\textwidth}
    \centering
    \includegraphics[width=\linewidth]{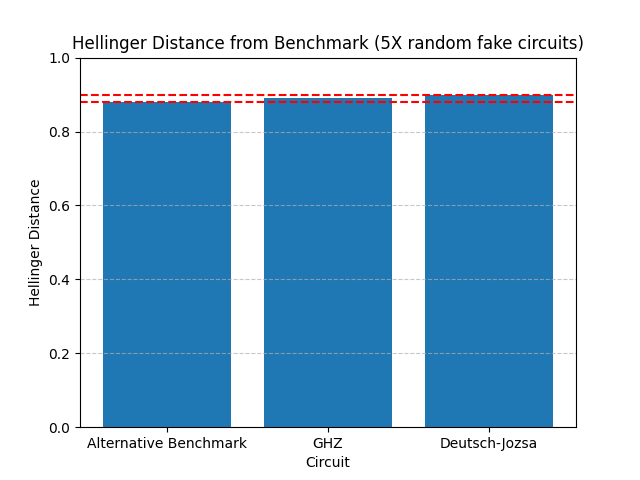}
    \caption{Random fake circuits}
    \label{fig:conf-rnd-5x-chart}
  \end{subfigure}\hfill
  \begin{subfigure}[b]{0.48\textwidth}
    \centering
    \includegraphics[width=\linewidth]{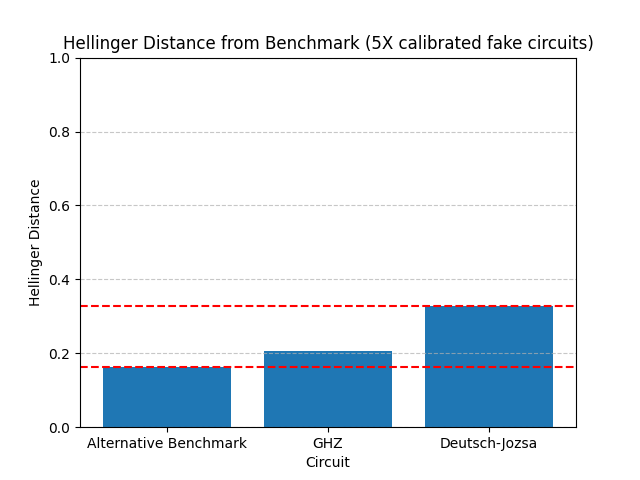}
    \caption{Calibrated fake circuits}
    \label{fig:conf-cal-5x-chart}
  \end{subfigure}
  \caption{Hellinger distance from benchmark circuit low-level \ac{QPU} output data with 5X fake circuits (lower is better, the red dotted lines delimit visually the range between the maximum and minimum value).}
  \Description{Hellinger distance from benchmark circuit low-level \ac{QPU} output data with 5X fake circuits (lower is better)}
  \label{fig:conf_5x_chart}
\end{figure}

Using five fake circuits for each real sub-circuit reduces the overall Hellinger distance between the \ac{QPUs} low-level outputs in the presence of different input circuits. However, as it can be seen in Table~\ref{table:hellinger-rnd-5x} and in Figure~\ref{fig:conf_5x_chart}, genuinely random fake circuits do not introduce anywhere near as much noise as calibrated ones.

\begin{table}[ht]
	\centering
	\begin{tabular}{|l|c|c|}
		\hline
		\textbf{Comparison Circuit} & \textbf{Hellinger Distance (random)} & \textbf{Hellinger Distance (calibrated)}\\[0.5ex]
		\hline
		Alternative benchmark & 0.880 & 0.163\\ \hline
		\ac{GHZ} & 0.891 & 0.206\\ \hline
		Deutsch-Jozsa & 0.898 & 0.328\\ \hline
	\end{tabular}
	\caption{Low-level output Hellinger distance from reference benchmark circuit with 5X random fake circuits.}
	\label{table:hellinger-rnd-5x}
\end{table}

\subsubsection{Third Test: 10X Fake Circuits}

\begin{figure}[htbp]
  \centering
  \begin{subfigure}[b]{0.48\textwidth}
    \centering
    \includegraphics[width=\linewidth]{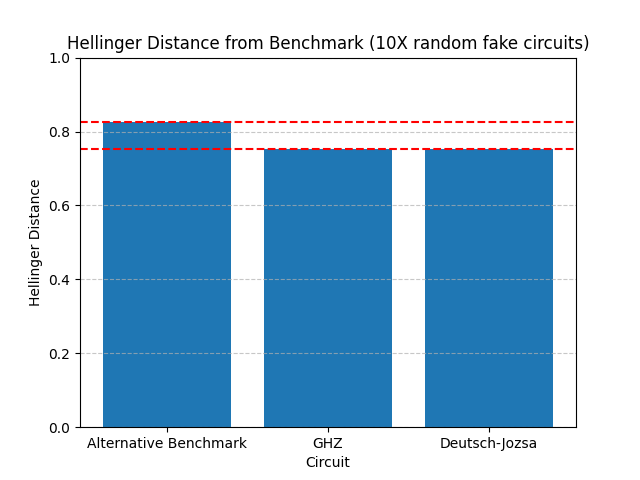}
    \caption{Random fake circuits}
    \label{fig:conf-rnd-10x-chart}
  \end{subfigure}\hfill
  \begin{subfigure}[b]{0.48\textwidth}
    \centering
    \includegraphics[width=\linewidth]{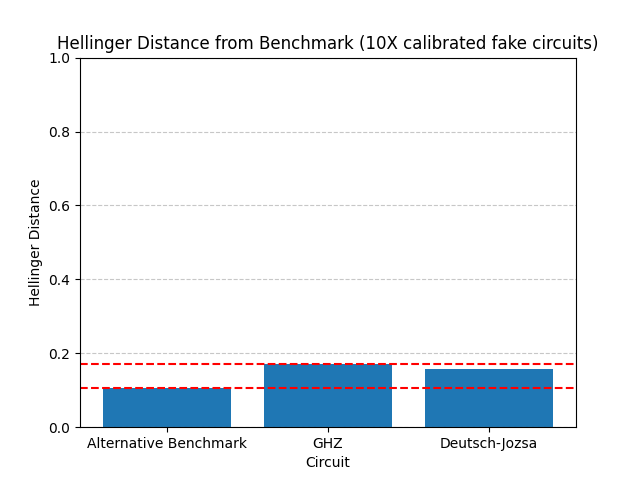}
    \caption{Calibrated fake circuits}
    \label{fig:conf-cal-10x-chart}
  \end{subfigure}
  \caption{Hellinger distance from benchmark circuit low-level \ac{QPU} output data with 10X fake circuits (lower is better, the red dotted lines delimit visually the range between the maximum and minimum value).}
  \Description{Hellinger distance from benchmark circuit low-level \ac{QPU} output data with 10X fake circuits (lower is better)}
  \label{fig:conf_10x_chart}
\end{figure}

In the final confidentiality-related test, for each sub-circuit, 10 fake circuits were also sent to each \ac{QPU}. The results, displayed in Table~\ref{table:hellinger-rnd-10x} and in Figure~\ref{fig:conf_10x_chart}, are even better than the ones with the 5X multiplier. As in the previous experiments, calibrated fake circuits remain the only option that provides sufficient noise to effectively mask differences between input circuits.

\begin{table}[ht]
	\centering
	\begin{tabular}{|l|c|c|}
		\hline
		\textbf{Comparison Circuit} & \textbf{Hellinger Distance (random)} & \textbf{Hellinger Distance (calibrated)}\\ [0.5ex]
		\hline
		Alternative benchmark & 0.826 & 0.107\\ \hline
		\ac{GHZ} & 0.753 & 0.172\\ \hline
		Deutsch-Jozsa & 0.742 & 0.157\\ \hline
	\end{tabular}
	\caption{Low-level output Hellinger distance from reference benchmark circuit with 10X random fake circuits}
	\label{table:hellinger-rnd-10x}
\end{table}

\subsubsection{Summary of the Results}
\label{subsec:discussion-confidentiality}

\begin{figure}[htbp]
  \centering
  \begin{subfigure}[b]{0.48\textwidth}
    \centering
    \includegraphics[width=\linewidth]{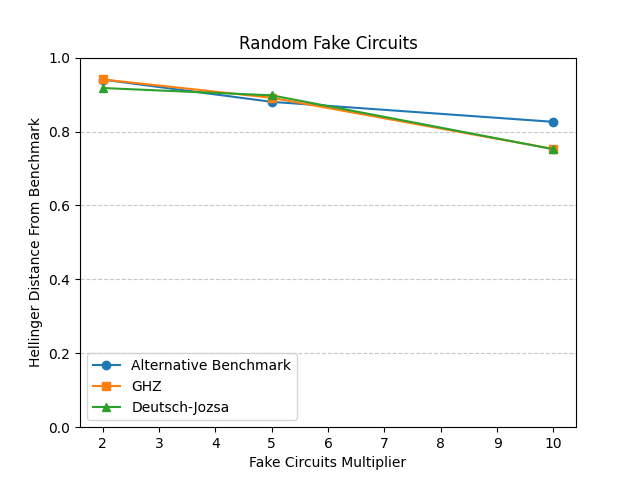}
    \caption{Random fake circuits}
    \Description{Random fake circuits}
    \label{fig:conf-rnd-recap}
  \end{subfigure}\hfill
  \begin{subfigure}[b]{0.48\textwidth}
    \centering
    \includegraphics[width=\linewidth]{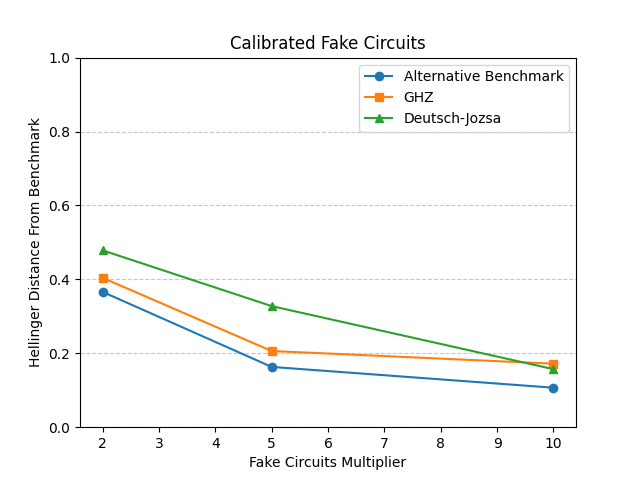}
    \caption{Calibrated fake circuits}
    \Description{Calibrated fake circuits}
    \label{fig:conf-cal-recap}
  \end{subfigure}
  \caption{Hellinger distance from benchmark circuit low-level \ac{QPU} output data with fake circuits (lower is better).}
  \Description{Hellinger distance from benchmark circuit low-level \ac{QPU} output data with fake circuits (lower is better)}
  \label{fig:conf_recap}
\end{figure}

Figure~\ref{fig:conf_recap} summarises the overall performance of truly random and calibrated fake circuits over different algorithms. With respect to the \ac{PNI}-inspired constraint defined in Section~\ref{sec:approach_overview}, which measures each \ac{QPU}'s ability to distinguish between different inputs. Note that the use of calibrated fake circuits yields superior performance compared to random fake circuits, given the same multiplier.

Wherever possible, calibrated sub-circuits should be the preferred countermeasure for obfuscating high-level, sensitive input from malicious \ac{QPUs}. In the general case, however, when it is not possible to estimate the magnitude and the range of a quantum circuit, truly random fake circuits can still offer a degree of confidentiality protection, even though they need higher multipliers with respect to calibrated ones. Additionally, it is important to note that circuit cutting inherently offers a level of input hiding by design, as long as no single \ac{QPU} receives the entire set of sub-circuits (see Subsection~\ref{subsubsec:probs2} for further discussion).

Overall, fulfilling \textbf{Goal G2}, as defined in Subsection~\ref{subsec:objectives}, is possible through the use of fake circuits in conjunction with circuit cutting, supported by the security countermeasures outlined in Subsection~\ref{subsec:countermeasures}.

\subsubsection{Proportional VS Exponential Sub-Circuit Allocation Probability, Revisited}
\label{subsubsec:probs2}

Subsection~\ref{subsubsec:probs2}, in the context of integrity-related security, summarised how a sub-circuit allocation probability model (which is exponential with respect to each \ac{QPU}'s integrity score) produced the best results. However, this came at the cost of underutilising legitimate \ac{QPUs} whose scores were only slightly lower than those of the best-performing ones, since saboteurs with very low scores were effectively ignored.

Note that, if we take into account the confidentiality of the input, allocating most of the work to a few or even a single \ac{QPU} poses significant security concerns. While the integrity-related attack surface is reduced, the confidentiality-related attack surface is extended. The proportional probability model guarantees a better input data distribution, which is obviously safer from a secrecy point of view.

Therefore, the proportional probability model for allocating sub-circuits to the \ac{QPUs} is the preferable choice if input data (the quantum circuit, which may be valuable \ac{IP}) confidentiality, is the asset to protect. In contrast, exponential probability is still the best choice when output data correctness is deemed more important.
\subsection{Experimental Methodology Generalisation}

The tests carried out to verify the resilience of various quantum circuit cutting schemes against integrity and confidentiality attacks represent not only a detailed validation of the proposed techniques, but it is also an example of the application of an experimental methodology guided by the principles of \ac{PNI}. This approach provides a conceptual framework that allows us to transform scenarios into quantitative metrics that, due to the nature of quantum computing and the specifics of security threats, are otherwise considerably difficult to evaluate objectively.

The adoption of a probabilistic framework such as \ac{PNI} allows the translation of security requirements (e.g., ``an attacker must not influence the global output'' or ``a processor must not distinguish between two different inputs'') into experimentally verifiable statistical relationships.

This approach is independent of the type of security technique under consideration and can therefore be reused to evaluate other protection strategies in distributed quantum systems. In other words, the methodology applied here for quantum circuit cutting constitutes a general model that can guide the design of experimental security property validations, providing a common framework for comparing different approaches and guiding system designer decisions.
\section{Threats to Validity}
\label{sec:threats}

In this section, we discuss potential threats to the validity of our findings, organised according to commonly accepted categories: internal, external, construct, and conclusion validity~\cite{wohlin2012experimentation}.

\subsection{Internal Validity}
Internal validity concerns whether the observed effects can be confidently attributed to the proposed countermeasures rather than to uncontrolled factors. Our evaluation is based on a software-based experimental test bed that combines Penny Lane's circuit-cutting facilities with noisy simulators. Malicious QPUs are modelled as devices that alter expectation values by a multiplicative factor of up to 250\%, and integrity scores are derived from probe circuits whose ground truth is known.

A first threat arises from the use of noisy QPU simulators rather than physical hardware. This choice makes our experimentation more reproducible but simplifies device-specific effects such as calibration drift, cross-talk, and correlated noise. Although the noise models employed (thermal relaxation and depolarising noise) approximate common NISQ behaviours, they do not capture the full variability of real devices. A second threat stems from our instantiation of integrity attacks as controlled perturbations to QPU outputs. These perturbations are intentionally strong but may not cover all realistic adversarial strategies. Finally, our integrity scores depend on the particular probe circuits used; alternative probes could yield different QPU rankings and, consequently, different sub-circuit allocation patterns.

\subsection{External Validity}
External validity concerns the extent to which our results generalise beyond the specific setting that we studied. The circuits evaluated (custom benchmarks, GHZ, Deutsch-Jozsa) provide structural diversity but do not represent the full spectrum of quantum workloads, particularly deeper or highly irregular algorithms. Our setup assumes six QPUs and 15-qubit circuits; larger systems, more QPUs, or heterogeneous hardware architectures may exhibit different behaviours, especially in terms of post-processing cost and fake-circuit overhead. Moreover, the brokered, non-communicating QPU architecture assumed in our threat model respects common current cloud models but may differ from highly custom cloud implementations, which could affect how adversarial capabilities materialise in practice.

\subsection{Construct Validity}
Construct validity refers to how well our operational measures capture the intended security concepts. We interpret \emph{integrity resilience} through the Hellinger distance between output distributions and adopt a fixed threshold (0.25) as an acceptable difference. While this choice is motivated by its relationship to total variation distance and Bayes error, it remains a heuristic; other thresholds or metrics (e.g., different statistical distances or task-specific utility measures) could lead to different conclusions.

Similarly, we approximate \emph{confidentiality resilience} by the ability of a QPU to distinguish between different high-level circuits given only its low-level outputs, with and without fake sub-circuits. This operationalisation focuses on statistical distinguishability and ignores other leakage channels, such as structural information in the sub-circuits, timing side-channels, or metadata about job submission. Our notion of calibrated fake circuits also assumes that the user may generate decoy workloads with expectation values similar in range and magnitude to the true sub-circuits. In scenarios where such knowledge is not available, the practical alignment between our experimental construct and the intended concept of input confidentiality would be weakened.

\subsection{Conclusion Validity}
Conclusion validity concerns whether the empirical evidence supports the claims drawn. Our analysis relies on repeated measurements of expectation-value distributions and on numerical comparison of Hellinger distances. Randomness in adversarial perturbations and fake-circuit generation, while controlled via seeding, may affect marginal cases near the similarity boundary. Furthermore, comparisons between proportional and exponential allocation strategies reflect average-case behaviours; atypical integrity-score distributions or outlier workloads could yield different outcomes.

In summary, our results are subject to the standard limitations of simulation-based evaluation and benchmark coverage, but we have taken multiple steps to promote fairness, reproducibility, and relevance. Future work will include validation on real quantum hardware and extended evaluations using additional circuit types and task-specific convergence objectives.
\section{Related Work}
\label{sec:related}

The research in the field of quantum-cloud security techniques is considerably dynamic. In this section we outline the state of the art with respect to techniques for the protection of integrity and confidentiality of quantum computations, and to verify their effectiveness.

\subsection{Integrity Protection}

Noise measurement and noise injection techniques, exploiting the physical error processes that naturally arise in current NISQ devices. By deliberately adding calibrated cross-talk or depolarising noise, these methods embed a fragile ``watermark’’ in the computation: any adversarial perturbation that exceeds the calibrated envelope can be statistically detected.  Recent works \cite{kundu2024qnad} shows that controlled cross-talk can improve the resistance of quantum‑mapped deep neural networks against gradient‑based attacks, increasing robustness on MNIST, CIFAR‑10, and CIFAR‑100 benchmarks by up to 268\%.  However, the approach remains limited to circuits whose logical structure tolerates additional noise.

\subsection{Confidentiality Protection}

Quantum circuit obfuscation is a family of confidentiality-preserving techniques that seek to conceal structural information gate types, data flow, and algorithmic intent, while preserving functional correctness. Two main families of obfuscation techniques dominate the literature: 
\begin{enumerate}
    \item \textbf{Dummy‐gate insertion}: strategically placed reversible gates (often CNOT) amplify the total variation distance (TVD) between the obfuscated and original output distributions, while barriers allow legitimate users to strip dummy entities after the circuit compilation \cite{suresh2021short}.
    \item \textbf{Circuit embedding}: random reversible sub‑circuits are interleaved with the target circuit, increasing the search space that an attacker must traverse \cite{das2023randomized}. 
\end{enumerate}

Both strategies increase the computation confidentiality, but they can conflict with compiler optimisations. Moreover, retaining obfuscation metadata without exposing it to a malicious tool-chain remains an open engineering challenge.

A second notable family of techniques for confidentiality protection is Blind quantum computation (BQC). BQC protocols guarantee information‑theoretic privacy by allowing a client to delegate quantum computation to an untrusted server without revealing information on input, algorithm, or output. The archetypal protocol by Broadbent \cite{broadbent2009universal} and its later refinements rely on the client ability to prepare single‑qubit states \(\lvert\!+\!_{\theta}\rangle\) with hidden phases and to engage in interactive, measurement‑based classical communication. While BQC may provide strong privacy and verifiability (via trap qubits), it assumes minimal quantum capabilities on the client side and incurs in substantial latency from round‑trip interactions. These requirements are incompatible with purely classical clients.

Finally, an interesting set of confidentiality-preserving techniques are Quantum homomorphic encryption (QHE) schemes.  QHE extends the classical notion of computation on encrypted data to the quantum domain.  Tan \cite{tan2016quantum} introduced a private‑key bosonic scheme in which encryption acts on internal degrees of freedom while computation acts on spatial modes, bounding the Holevo information by \(\chi\le m\log_2 m\). Follow‑up photonic prototypes demonstrated the principle experimentally but they remain limited to small‑dimensional Hilbert spaces and non‑interactive gate sets. Fundamental obstacles currently restrict QHE to proof-of-concept demonstrations: the no‑cloning theorem, decoherence under ciphertext growth, and polynomial overheads for universal gate sets.

\subsection{Noninterference-based Quantum Circuit Security Validation}

Recently, Ying et al. proposed an extension of the classical \ac{PNI} model for the formal verification of the security properties of quantum algorithms \cite{6595825}. In the quantum context, the presence of entanglement and the inevitable state alteration due to measurements introduce new challenges, making classical \ac{PNI} formulations indirectly inapplicable. To address these peculiarities, a quantum automaton model has been proposed that uses density operators, super-operators, and POVM measurements to describe the evolution of the system and observations of the agents. On this basis, noninterference is reformulated in terms of the insecurity degree, which quantifies the extent of possible interference between groups of agents, and an approximate version is introduced to account for noise and physical imperfections. The approach is supported by a generalisation of the unwinding technique for formal verification, which allows for the estimation of upper bounds on the insecurity degree, and by a compositionality theorem that ensures the preservation of security in composite systems, in the absence of entanglement between components.

\subsection{Positioning of This Work}

Existing integrity checks, such as noise injection, are algorithm‑specific; circuit obfuscation alone does not ensure correctness; BQC mandates a quantum‑capable client, and QHE is still impractical for medium‑scale circuits.

Our paper addresses these gaps by proposing a generic circuit cutting-based approach that can be easily implemented and integrated in existing quantum-cloud infrastructures. Unlike prior art, our method scales to any number of logical qubits without client‑side quantum resources, bridging the gap between near‑term NISQ capabilities and cryptographic‑grade security requirements both in terms of information integrity and confidentiality.
\section{Conclusions}
\label{sec:conclusions}

In this paper, we discussed whether quantum circuit cutting (when properly enhanced) can serve not only as a circuit decomposition technique to make use of \ac{QPUs} with a limited number of qubits, but also as an effective security mechanism. In particular, we aimed to explore and outline a set of security countermeasures that, when combined with circuit cutting, can enhance the integrity and confidentiality of quantum computations in potentially untrusted or adversarial environments.

As outlined in Subsection~\ref{subsec:objectives}, the objectives of this works are as follows:

\begin{description}
	\item[Goal G1:] Proposing a methodology to improve the \textit{integrity} of output data while using potentially untrusted quantum processors through a trusted centralised scheduler.
	
	\item[Goal G2:] Proposing a methodology to improve the \textit{confidentiality} of algorithms and input/output data while using potentially untrusted quantum processors through a trusted centralised scheduler.
    
    \item[Goal G3:] Designing a methodological framework for quantum computational system designers to compare different quantum architectures with respect to integrity and confidentiality resilience.
\end{description}

To address both goals, \textbf{G1} and \textbf{G2}, the key idea was to use enhanced quantum circuit cutting. In light of the experimental results presented in Section~\ref{sec:exps}, we now revisit the contributions of this work with respect to the previously stated goals.

\subsubsection{Goal G1}

The resilience of quantum circuit cutting with respect to integrity has been analysed within the guidelines of the Probabilistic Noninterference (PNI) framework under several configurations:

\begin{enumerate}
	\item raw quantum circuit cutting;
	\item circuit cutting enhanced with dynamic \ac{QPU} integrity scores, sub-circuit distribution with a probability proportional to such scores, and a 2X sub-circuit replication factor with weighted averaging of \ac{QPU} results;
	\item same as above, but with a probability exponential with respect to integrity scores and a 2X sub-circuit replication;
	\item same as above, but with a 3X sub-circuit replication;
	\item same as above, but without sub-circuit replication.
\end{enumerate}

The last configuration is the one that yields the best results:

\begin{itemize}
	\item dynamic integrity scores computed with probe circuits evaluation by each \ac{QPU};
	\item sub-circuit allocation probabilities for each \ac{QPU} exponential with respect to integrity scores;
	\item no sub-circuit replication.
\end{itemize}

This configuration yields the best results even with 5 attacking \ac{QPUs} over 6. The technique can be easily implemented and integrated within existing shot-wise quantum broker architectures such as \cite{fi17110507,10.1007/978-3-031-48421-6_25} and \cite{10313701}.

\subsubsection{Goal G2}

The \ac{PNI}-inspired requirement for confidentiality is that the \ac{QPUs} should not be able to distinguish between different input circuits. Raw circuit cutting, while offering a degree of secrecy for both input and output data by dividing the quantum circuit into multiple sub-circuits allocated to different \ac{QPUs}, falls short of meeting this more stringent requirement.

Fake circuits were mixed with real sub-circuits obtained by quantum circuit cutting to hide input circuit differences to the \ac{QPUs}. Calibrated fake circuits\footnote{Fake circuits whose expectation values fall within the range of the actual sub-circuits expectation values and are of a comparable magnitude.} with a specific multiplier (e.g., five fake circuits for each real sub-circuit) yielded the best results. With a 5X or greater multiplier, input circuit differences are effectively hidden to the individual \ac{QPUs}. In the general case, where no information is available about the characteristics of the input circuit's expectation values, random fake sub-circuits can still be used. However, it should be noted that they require a larger multiplier than calibrated fake circuits to achieve a comparable level of confidentiality protection.

We tested several multipliers  (2X, 5X, and 10X) with different algorithms (custom benchmark circuits, \ac{GHZ} and Deutsch-Jozsa), and the overall trend is that higher fake circuit multipliers provide greater input data obfuscation. The optimal choice of multiplier depends on the trade-off between the relevance assigned to input data confidentiality and the cost associated with processing fake data.

Overall, quantum circuit cutting enhanced with fake circuit flooding (calibrated fake circuits, when possible) provides an adequate confidentiality protection for the input data of quantum computations. Note that these countermeasures can also be easily integrated into existing shot-wise quantum broker architectures such as \cite{fi17110507,10.1007/978-3-031-48421-6_25} and \cite{10313701}.

\subsubsection{Goal G3}

The \ac{PNI}-inspired heuristics have proven themselves effective in terms of system design explorations:

\begin{enumerate}
    \item Estimating the integrity resilience of a specific architecture can be performed by means of \ac{QPU} simulators, whose outputs will be stored firstly unaltered, then corrupted by an increasing number of malicious \ac{QPUs}. The attacker capabilities can be defined flexibly, depending on the context, and the results are obtained in a relatively short amount of time.
    \item Estimating confidentiality resilience requires some care with respect to the input provided to the system: truly random inputs will hardly satisfy the \ac{PNI}-inspired constraints. Therefore, calibrated inputs that produce outputs close to the ground truth are the preferable option.
\end{enumerate}

The heuristics defined in this paper can be readily integrated into quantum software engineering workflows, as they require relatively little time to execute and provide system designers with valuable guidance on specific security properties. However, as shown in Subsection~\ref{subsubsec:probs2}, a rigorous assessment of trade-offs is still needed, since different security properties may lead to conflicting requirements. It is therefore the responsibility of the system designer to reconcile these conflicts while keeping the overall system objectives in mind. This reflects a broader principle in software engineering: even in the case of quantum computing, security must be treated as a first-class design concern, carefully balanced against performance, usability, and other non-functional requirements throughout the development life-cycle.

\subsection{Secure Quantum Scheduler}
\label{subsec:secure-quantum-scheduler}

A notable application of the security techniques proposed in this work is the design of a secure scheduler within the cloud-based quantum computing scenario outlined in Subsection~\ref{subsec:scenario}. Here, we reflect on this issue and discuss its implications for system architecture and trust assumptions.

The secure scheduler should protect the user's quantum circuit (the input) and its computation results (the output) both from a confidentiality and integrity point of view. As stated in Subsection~\ref{subsubsec:probs2}, there are trade-offs to be made depending on whether input data confide~ntiality or output data integrity are deemed more important. In real-world contexts, users often operate under different constraints, and a secure system should enable them to make the best choices based on their specific needs.

To simplify the end user's decision-making, the secure scheduler can offer three macro-configurations based on the user's security preferences. Each configuration combines quantum circuit cutting with the additional countermeasures defined in Subsection~\ref{subsec:countermeasures}\footnote{Please note that sub-circuit replication is not included, since the experiments of Subsection~\ref{subsec:integrity} proved it to be detrimental to integrity resilience and, intuitively, distributing the same sub-circuit to more than one \ac{QPU} broadens the confidentiality attack surface as well.}. A hypothetical scheduler user could specify the desired security profile along with other computation-related parameters through a user interface or a manifest file, possibly fine-tuning specific parameters if needed. Note that the following configurations are only meant to show the interplay between the findings related to \textbf{Goal G1} (integrity) and \textbf{Goal G2} (confidentiality); the specific set of security countermeasures and their parameters will have to undergo experimental verification before being usable in security-critical contexts.

The three configurations presented below represent different balance points between two often competing objectives: protecting input confidentiality and protecting output integrity. The choice of a specific profile depends on the operational context and the type of risk perceived by the user.

\subsubsection*{Configuration 1 - Confidentiality Over Integrity:}
\begin{itemize}
	\item Fake sub-circuits with a 10X factor compared to the real ones (the user can optionally fine-tune the factor).
	\item Integrity probe circuits to compute integrity scores.
	\item Sub-circuit allocation probabilities \textit{proportional} to the sum of the confidentiality score $CS$ multiplied by two and the \ac{QPUs} dynamic integrity score $IS$:
	\begin{equation}
		\dfrac{2CS_{n} + IS_{n}}{\sum_{i=1}^{n\_qpus} 2CS_{i} + IS_{i}},
	\end{equation}
	where $n$ is the current \ac{QPU} index and $n\_qpus$ is the number of available \ac{QPUs}.
\end{itemize}

This configuration prioritises confidentiality, assuming scenarios where unauthorised access to the quantum circuit is the primary threat (e.g., multi-tenant environments where untrusted operators manage the processors). Using a high factor of fake circuits reduces the ability of malicious \ac{QPUs} to distinguish the real circuit.

\subsubsection*{Configuration 2 - Balanced:}
\begin{itemize}
	\item Fake sub-circuits with a 5X factor compared to the real ones (the user can optionally fine-tune the factor).
	\item Integrity probe circuits to compute integrity scores.
	\item Sub-circuit allocation probabilities \textit{proportional} to the sum of the \ac{QPUs} dynamic integrity score $IS$ and the confidentiality score $CS$:
	\begin{equation}
		\dfrac{CS_{n} + IS_{n}}{\sum_{i=1}^{n\_qpus} CS_{i} + IS_{i}},
	\end{equation}
 where $n$ is the current \ac{QPU} index and $n\_qpus$ is the number of available \ac{QPUs}.
\end{itemize}

This configuration is designed for scenarios where both confidentiality and integrity are important and the risk is distributed. The false circuit factor and the equal weighting between confidentiality and integrity scores aim to provide balanced protection, reducing overall risk without optimising one objective at the expense of the other. It is suitable for contexts where it is not possible to determine a priori which threat is predominant.

\subsubsection*{Configuration 3 - Integrity over Confidentiality:}
\begin{itemize}
	\item Fake sub-circuits with a 2X factor compared to the real ones (the user can optionally fine-tune the factor).
	\item Integrity probe circuits to compute integrity scores.
	\item Sub-circuit allocation probabilities \textit{exponential} with respect to the sum of \ac{QPUs} dynamic integrity score $IS$ multiplied by two and the confidentiality score $CS$:
	\begin{equation}
		\dfrac{e^{2IS_{n} + CS_{n}}}{\sum_{i=0}^{n\_qpus} e^{2IS_{i} + CS_{i}}},
	\end{equation}
	where $n$ is the current \ac{QPU} index and $n\_qpus$ is the number of available \ac{QPUs}.
\end{itemize}

This configuration is suitable when the accuracy of the final result is a priority, such as critical applications where altered output could cause serious consequences. The reduced number of falsified circuits limits the overhead and maintains a high quality of the data used for output reconstruction. Allocation probabilities strongly favour \ac{QPUs} with high integrity scores, making it more difficult for an attacker to significantly alter the overall result.

\subsection{Future Work}

To conclude, we outline several promising research directions that could further develop the use of quantum circuit cutting as a security technique for enhancing integrity and confidentiality.

\begin{description}
	\item[Delayed integrity probes:] rather than sending integrity probes upfront, before the actual quantum computation, they could be mixed at any later point of sub-circuit execution, be it the middle or even the end; by doing this, especially in combination with fake circuits, it would become more challenging for malicious \ac{QPUs} to detect and avoid the probes, making the integrity score more reliable.
	\item[Improved integrity probes:] noise injection and measurement techniques could be used instead of or together with circuits with known expectation values;
	\item[Sub-circuits as integrity probes:] replicated sub-circuit evaluation could be used as an integrity probe if most of the \ac{QPUs} are considered reliable.
	\item[Integration with alternative security techniques:] security techniques such as Circuit Obfuscation (see Section~\ref{sec:related}) could be integrated with quantum circuit cutting as further security enhancements and evaluated through the \ac{PNI}-based methodology developed in this work.
	\item[Secure Quantum Scheduler:] explore the integration of integrity and confidentiality-related security countermeasures in a real-world context to provide end-users with a robust and usable system (Subsection~\ref{subsec:secure-quantum-scheduler} sketches a hypothetical direction for this line of research).
\end{description}

\begin{acks}
This work has been partially funded by SEcurity and Rights in the CyberSpace - SERICS (PE00000014 - CUP H73C2200089001) under the National Recovery and Resilience Plan funded by the European Union - NextGenerationEU.
\end{acks}

\bibliographystyle{ACM-Reference-Format}
\bibliography{biblio}


\end{document}